\documentclass[sn-mathphys,Numbered]{sn-jnl}
\usepackage{graphicx}%
\usepackage{multirow}%
\usepackage{amsmath,amssymb,amsfonts,mathabx}%
\usepackage{amsthm}%
\usepackage{mathrsfs}%
\usepackage[title]{appendix}%
\usepackage{xcolor}%
\usepackage{textcomp}%
\usepackage{manyfoot}%
\usepackage{booktabs}%
\usepackage{algorithm}%
\usepackage{algorithmicx}%
\usepackage{algpseudocode}%
\usepackage{listings}%
\usepackage{nicematrix}
\usepackage[makeroom]{cancel}
\usepackage{imakeidx}
\usepackage{tikz}
\usepackage{soul}
\usepackage{overpic}

\usepackage{algorithm}
\usepackage{algpseudocode}

\usetikzlibrary{tikzmark}
\usetikzlibrary{calc}
\usetikzlibrary{arrows.meta}

\definecolor{matblue}{rgb}{0.000 0.447 0.741}
\definecolor{matred}{rgb}{0.850 0.325 0.098}
\definecolor{matyellow}{rgb}{0.9290 0.6940 0.125}
\definecolor{matpurple}{rgb}{0.494 0.184 0.556}
\definecolor{matgreen}{rgb}{0.466 0.674 0.188}
\definecolor{green}{rgb}{0.466 0.674 0.188}

\providecommand\bnabla{\boldsymbol{\nabla}} 

\begin{document}

\title[An Invitation to Resolvent Analysis]{An Invitation to Resolvent Analysis}

\author*[1]{\fnm{Laura Victoria} \sur{Rolandi}}\email{vrolandi@ucla.edu}
\author[2]{\fnm{Jean Hélder} \sur{Marques Ribeiro}}\email{jean.marques@fem.unicamp.br}
\author[3]{\fnm{Chi-An} \sur{Yeh}}\email{chian.yeh@ncsu.edu}
\author[1]{\fnm{Kunihiko} \sur{Taira}}\email{ktaira@seas.ucla.edu}

\affil[1]{\orgdiv{Department of Mechanical and Aerospace Engineering}, \orgname{University of California, Los Angeles, CA 90095, USA}}
\affil[2]{\orgdiv{Faculdade de Engenharia Mecânica}, \orgname{Universidade Estadual de Campinas, Brazil}}
\affil[3]{\orgdiv{Department of Mechanical and Aerospace Engineering}, \orgname{North Carolina State University, Raleigh, NC 27695, USA}}

\abstract{
Resolvent analysis is a powerful tool that can reveal the linear amplification mechanisms between the forcing inputs and the response outputs about a base flow.  These mechanisms can be revealed in terms of a pair of forcing and response modes and the associated gains (amplification magnitude) in the order of energy contents at a given frequency.  The linear relationship that ties the forcing and the response is represented through the resolvent operator (transfer function), which is constructed through spatially discretizing the linearized Navier-Stokes operator.  One of the unique strengths of resolvent analysis is its ability to analyze statistically stationary turbulent flows.  In light of the increasing interest in using resolvent analysis to study a variety of flows, we offer this guide in hopes of removing the hurdle for students and researchers to initiate the development of a resolvent analysis code and its applications to their problems of interest.  To achieve this goal, we discuss various aspects of resolvent analysis and its role in identifying dominant flow structures about the base flow.  The discussion in this paper revolves around the compressible Navier-Stokes equations in the most general manner. We cover essential considerations ranging from selecting the base flow and appropriate energy norms to the intricacies of constructing the linear operator and performing eigenvalue and singular value decompositions.  Throughout the paper, we offer details and know-how that may not be available to readers in a collective manner elsewhere.  Towards the end of this paper, examples are offered to demonstrate the practical applicability of resolvent analysis, aiming to guide readers through its implementation and inspire further extensions.  We invite readers to consider resolvent analysis as a companion for their research endeavors.
}


\maketitle

\tableofcontents

\section{Introduction}
\label{sec1}

In turbulence, unsteady flow structures across a range of spatiotemporal scales emerge, evolve, and interact in a complex manner giving rise to its vibrantly rich dynamics.  Understanding the fundamental mechanisms of unsteady and turbulent flows has been of critical importance in fluid dynamics \cite{reynolds1883xxix, rayleigh1878instability, rayleigh1879stability, huerre1990local, Schmid01}.  To study these flows, mathematically rigorous and systematic characterization of the spatiotemporal evolution of flow perturbations is key in not only deepening our insights but also for modifying the dynamics of such flows to achieve engineering benefits, including drag reduction, lift increase, noise mitigation, mixing improvement, and combustion enhancement.  

Because fluid flows generally require high degrees of freedom to describe their dynamics even for relatively simple problems, tracking and understanding all unsteady flow structures remains a daunting task.  Therefore, what is often pursued is to study the dominant dynamics that offer accurate predictions of the primary dynamics (large-scale motions) or turbulent statistics.  
Capturing the dominant behavior of the flow and the associated spatial structures can be achieved with modal analysis \cite{holmes2012turbulence, SchmidJFM2010dmd, Kutz2016book, Theofilis:ARFM11, Schmid01, Jovanovic:ARFM21, taira2017modal, taira2020modal, unnikrishnan2023recent}.  At its core, modal analysis methods seek to break down a flow field into a linear combination of modes. These modes can capture the dominant dynamics of the system, enabling analysis, modeling, estimation, and control \cite{holmes2012turbulence, Aubry:JFM88, Rowley:ARFM17, Noack2003jfm, Noack11, brunton2015closed}. 

These modal analysis techniques can be broadly categorized into data-based and operator-based techniques.  Data-based techniques include the proper orthogonal decomposition \cite{holmes2012turbulence, Sirovich1987} and dynamic mode decomposition \cite{SchmidJFM2010dmd, Kutz2016book}.  Operator-based techniques are based on the linearized Navier--Stokes equations and include the global stability analysis \cite{Theofilis:ARFM11,Theofilis:PAS03} and resolvent analysis \cite{trefethen1993hydrodynamic, Farrell:JAS1996, jovanovic2005componentwise, mckeon2010critical}.  In fact, the global stability analysis and resolvent analysis are closely related to one another \cite{taira2017modal}.  Given the linearized governing equation, the global stability analysis examines the stability of the base flow through eigenvalue decomposition and captures the homogeneous solution to the initial value problem.  Resolvent analysis on the other hand considers the same but forced linear system, which can be examined through singular value decomposition.  In a complementary manner, the resolvent framework studies the particular solution to the forced problem.  For these reasons, it is known that the global stability analysis and resolvent analysis go hand in hand to provide a complete picture of the linear flow dynamics.  

Resolvent analysis reveals the amplification dynamics for a given base flow through pairs of forcing and response modes and the gain that signifies the magnitude of amplification \cite{trefethen1993hydrodynamic,jovanovic2005componentwise}.  Initially, resolvent analysis was formulated for studying stable laminar flows (solutions to the Navier--Stokes equations).  This formulation was later extended to unstable systems \cite{jovanovic2004modeling} and the analysis of time-averaged turbulent flows \cite{mckeon2010critical}. The latter extends from the former approach by considering the nonlinear terms as a self-sustained input within the natural feedback mechanism of the flow system.  The resolvent analysis essentially analyzes the pseudospectra of the linear operator, revealing harmonic response characteristics and the transient energy growth \cite{trefethen1993hydrodynamic, jovanovic2005componentwise}.  These perspectives have greatly expanded the horizon of modal analysis to examine complex turbulent flows \cite{taira2017modal,taira2020modal}. 

Resolvent analysis has been applied to study diverse types of flow problems, including channel flows \cite{jovanovic2005componentwise, moarref2013model}, pipe flows \cite{mckeon2010critical}, boundary layers \cite{nogueira2020resolvent}, wing wakes \cite{thomareis2018resolvent,yeh2020resolvent, ribeiro2023triglobal}, turbulent jets \cite{Jeun:PF16input,schmidt2018spectral}, combustion \cite{skene2019adjoint}, airfoil noise \cite{ricciardi2022transition}, transonic buffet \cite{kojima2020resolvent,Houtman:Flow23resolvent} and cavity flows \cite{sun2020resolvent,liu2021unsteady}.  Because resolvent analysis is essentially equivalent to the state-space system description from the dynamical system and control theories, it naturally provides guidance on designing active and passive flow control strategies \cite{luhar2014opposition, toedtli2019predicting, yeh2019resolvent, liu2021unsteady, lin2023flow, gross2024laminar, ribeiro2024control}.  It can also serve as the basis to perform flow estimation \cite{Towne:JFM20}. Therefore, resolvent analysis represents a valuable complement to both flow simulations and experiments by offering insights into the fundamental understanding and modeling of fluid flows. 

With growing interest in utilizing resolvent analysis to understand the input-output relation for a variety of flows, many research groups have initiated efforts to develop and implement computer programs to perform resolvent analysis.  During their efforts, our research groups have been asked on many occasions how to handle some of the details and nuances related to their code developments and how to interpret their results.  Some of these details and know-how are not necessarily discussed in archival papers and are not made available in a single document.  Here, we attempt to provide as many of these details as possible in a single guide to assist those who may be considering adopting resolvent analysis as their tool of choice.  For this reason, this document is not designed to serve as a review article.  For a survey on resolvent analysis, we point to \cite{Jovanovic:ARFM21}.

\begin{figure}
\centering
\includegraphics[width=\textwidth]{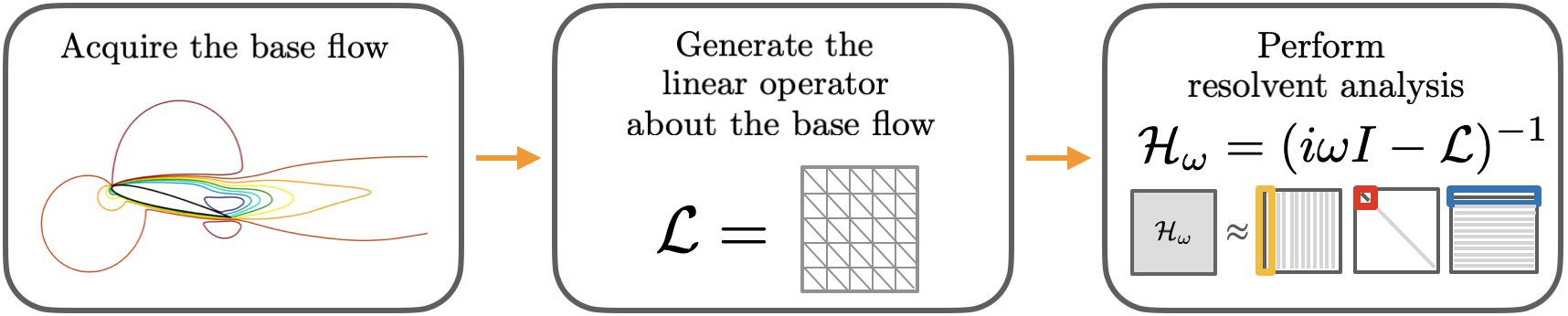}
\caption{\label{fig:MainStepDiagramResolvent}Overview of performing resolvent analysis.} 
\end{figure}

This document aims to serve as a comprehensive guide for implementing resolvent analysis, with detailed discussions and suggestions based on the authors’ use of resolvent analysis over the years.  Unlike other papers, we discuss practical matters that should be considered for the construction of the linear operator and numerical procedures to compute the resolvent modes.  The basic steps for resolvent analysis are summarized in Fig. \ref{fig:MainStepDiagramResolvent}, which forms the basis of this paper.  To offer concrete recommendations for using resolvent analysis in this paper, we mainly focus on the compressible Navier--Stokes equations.  

The structure of this paper is as follows.  The basic concept of resolvent analysis and its implementation are discussed in Sect. \ref{sec:Resolvent}.   Computational approaches to find the resolvent modes and gain are offered in Sect. \ref{sec:Methods}.  Finally, in Sect. \ref{sec:StepByStep}, we detail the setup for performing the resolvent analysis and provide the physical interpretation of the results through biglobal and triglobal resolvent analysis examples.  Concluding remarks are offered in Sect.~\ref{sec:Conclu}.  In addition to the main text, supplemental material on the full linearized compressible Navier--Stokes equations and pseudocodes for performing the resolvent analysis are made available in the Appendix.  We hope this document serves as a welcoming invitation to new students and researchers interested in resolvent analysis.

\section{Resolvent analysis}
\label{sec:Resolvent}
In this section, we present the formalism of resolvent analysis, provide considerations about the appropriate energy norm and discuss the selection of the base flow, which is a critical aspect for meaningful analysis. We also provide guidance on generating the linear operator and introduce variations from the classical approach that can enhance the effectiveness of resolvent analysis. 

\subsection{Formulation}
\label{sec:inputoutput}

Let us consider the Navier--Stokes (NS) equations for three-dimensional compressible flow\footnote{We present resolvent analysis for compressible flow in this paper.  For incompressible flow, see \cite{mckeon19notes,mckeon2010critical} for details.}, expressed with conservative variables:
\begin{equation}
    \begin{split}
        &\frac{\partial \rho}{\partial t} + \bnabla \cdot ( \rho \mathbf{u}) = g_\rho \\
        &\frac{\partial  \rho \mathbf{u}}{\partial t} + \bnabla \cdot ( \rho \mathbf{u}\otimes \mathbf{u}) =-\bnabla p+\bnabla \cdot \left[\mu\;(\bnabla\mathbf{u}+\bnabla\mathbf{u}^T)-\frac{2}{3}\mu\bnabla\cdot\mathbf{u}\:\mathrm{I}\right] + \mathbf{g}_{\rho\mathbf{u}}\\
        &\frac{\partial  \rho e}{\partial t}+\bnabla \cdot ( \rho \mathbf{u}e+p\mathbf{u})=\bnabla \cdot (K\bnabla T)+\bnabla \cdot \left(\mathbf{u}\;\left[\mu\;(\bnabla\mathbf{u}+\bnabla\mathbf{u}^T)-\frac{2}{3}\mu\bnabla\cdot\mathbf{u}\:\mathrm{I}\right]\right) + g_{\rho e},
    \end{split}
    \label{eq:NS-cons}
\end{equation}
where $\rho$ is the fluid density, $\mathbf{u}=(u_x,u_y,u_z)$ the velocity vector, $e$ is the total energy, $p$ the pressure, and $T$ the temperature. 
In these equations, external source and forcing terms are present in each equation represented by $g_\rho$, $\mathbf{g}_{\rho \mathbf{u}}$, and $g_{\rho e}$.
The viscosity $\mu$ and the thermal conductivity $K$ are taken to be constant. 
The above set of equations, Eqs.~\ref{eq:NS-cons}, can be spatially discretized on a grid using a computational fluid dynamics technique of one's choice.  We can then express the discretized NS equations in compact form
\begin{equation}\label{eq:NSCompact}
    \frac{d\mathbf{q}}{dt}=\mathcal{N}(\boldsymbol{\mathbf{q}}) + \mathbf{g},
\end{equation}
where $\mathbf{q}=(\rho, \rho \mathbf{u},\rho e)^T \in\mathbb{R}^N$ represents the state vector (conservative variables) and $\mathcal{N}$ denotes the nonlinear NS evolution operator.  The degrees of freedom of the state variable $N$ depend on the spatial discretization. If $N_{\text{cell}}$ indicates the number of cells/points used for the spatial discretization and $m$ is the number of state variables, then $N=m\times N_{\text{cell}}$. The last term $\mathbf{g} = (g_\rho, \mathbf{g}_{\rho \mathbf{u}}, g_{\rho e})^T$ collects the external forcing terms.  Because Eq.~\ref{eq:NSCompact} is taken to be spatially discretized, the temporal derivative is expressed with an ordinary derivative instead of a partial derivative.

Now, we consider the flow field to be comprised of the chosen time-invariant base flow $\mathbf{q}_b=(\bar{\rho},\widebar{\rho \mathbf{u}},\widebar{\rho e})^T \in \mathbb{R}^N$ and the perturbation $\mathbf{q}^\prime=(\rho^\prime,(\rho \mathbf{u})^\prime,(\rho e)^\prime)^T \in \mathbb{R}^N$ 
such that 
\begin{equation}
    \mathbf{q}(\mathbf{x},t) = \mathbf{q}_b(\mathbf{x}) 
                             + \mathbf{q}^\prime(\mathbf{x},t).
\end{equation} 
Substituting this velocity field expression into Eq.~\ref{eq:NSCompact} and performing a Taylor series expansion, we obtain
\begin{equation}
        \frac{d \mathbf{q}_b}{d t}+\frac{d \mathbf{q}^\prime}{d t}
        =\mathcal{N}(\mathbf{q}_b+\mathbf{q}^\prime) + \mathbf{g}
        =\mathcal{N}(\mathbf{q}_b)
        +\left. \bnabla_{\mathbf{q}} \mathcal{N} \right|_{\mathbf{q}_b}\mathbf{q}^\prime+\mathbf{g}+O(|\mathbf{q}^\prime|^2).
\end{equation}
Because we are considering a stationary base flow (${d \mathbf{q}_b}/{d t}=0$), we arrive at the system describing the dynamics of the perturbation $\mathbf{q}^\prime$ as
\begin{equation}\label{eq:lin}
\frac{d \mathbf{q}^\prime}{d t}=\mathcal{L}\mathbf{q}^\prime +\mathbf{f}^\prime.
\end{equation}
In this equation, 
\begin{equation}
    \mathcal{L}\equiv \left. \bnabla_{\mathbf{q}} \mathcal{N} \right|_{\mathbf{q}_b}
    \in\mathbb{R}^{N\times N}
    \label{eq:linear_jacobian}
\end{equation}
represents the linearized NS operator about the base flow $\mathbf{q}_b$ (see Appendix \ref{AppendixA} for its continuous form) while 
\begin{equation}\label{eq:forcing}
    \mathbf{f}^\prime
    =
    \mathcal{N}(\mathbf{q}_b)
    +O(|\mathbf{q}^\prime|^2) + \mathbf{g}\in\mathbb{R}^{N}
\end{equation}
collects the nonlinear NS operator acting on the base flow, the nonlinear terms, and external forcing $\mathbf{g}$.  If the problem has a periodic direction as in the case of biglobal analysis, we can have a complex variable formulation.  See Appendix \ref{AppendixA} for details.
Equation \ref{eq:lin} represents the evolution of the response $\mathbf{q}^\prime$ through the linearized NS equations with forcing $\mathbf{f}^\prime$.

Through resolvent analysis, we seek the dominant forcing inputs and the corresponding perturbations that are amplified over the flow field. 
Given a stationary base flow, we assume the perturbations to be harmonic (sinusoidal in time).  This means that the response ($\mathbf{q}^\prime$) and forcing ($\mathbf{f}^\prime$) can be expressed through Fourier representation of
\begin{equation}
\begin{split}
    \mathbf{q}^\prime(\mathbf{x,t})=\int^\infty_{-\infty}\hat{\mathbf{q}}_\omega(\mathbf{x})e^{-i\omega t}d\omega\\
    \mathbf{f}^\prime(\mathbf{x,t})=\int^\infty_{-\infty}\hat{\mathbf{f}}_\omega(\mathbf{x})e^{-i\omega t}d\omega.
\end{split}
\label{eq:FourierTrans}
\end{equation}
With these Fourier representations, Eq.~(\ref{eq:lin}) can be rewritten as
\begin{equation}
    - i\omega\hat{\mathbf{q}}_\omega(\mathbf{x})
    =\mathcal{L}\hat{\mathbf{q}}_\omega(\mathbf{x})
    +\hat{\mathbf{f}}_\omega(\mathbf{x})
    \label{eq:freq_space_eq}
\end{equation}
at each temporal frequency $\omega$, providing the following input-output relationship:
\begin{equation}
    \hat{\mathbf{q}}_\omega(\mathbf{x})
    =
    [-i\omega I -\mathcal{L}]^{-1}\hat{\mathbf{f}}_\omega(\mathbf{x})
    =\mathbf{\mathcal{H}}_\omega\hat{\mathbf{f}}_\omega(\mathbf{x}).
    \label{eq:InputOutput}
\end{equation}
Here, $\mathbf{\mathcal{H}}_\omega=[-i\omega I-\mathcal{L}]^{-1}\in \mathbb{C}^{N\times N}$ is the {\it resolvent operator} that acts as a transfer function between the forcing input and the response output at frequency $\omega$. 

In resolvent analysis, instead of examining all possible combinations of $ \hat{\mathbf{q}}_\omega(\mathbf{x})$ and $ \hat{\mathbf{f}}_\omega(\mathbf{x})$, our objective is to identify the optimal ones that maximize the energy between inputs and outputs of the system \citep{schmid2007nonmodal}
\begin{equation}
    \sigma_\omega^2 
    =\max_{\hat{\mathbf{f}}_\omega}
    \frac{\langle\hat{\mathbf{q}}_\omega,~\hat{\mathbf{q}}_\omega\rangle_{E}}{\langle\hat{\mathbf{f}}_\omega,~\hat{\mathbf{f}}_\omega\rangle_{E}}
    =\max_{\hat{\mathbf{f}}_\omega}
    \frac{||\hat{\mathbf{q}}_\omega||_{E}}{||\hat{\mathbf{f}}_\omega||_{E}},
    \label{eq:optmization}
\end{equation}
where $\sigma_\omega$ is the energy gain and $||\cdot||_{E}$ is a suitable energy norm.  This norm needs to be chosen as a physically relevant metric, which we will discuss in detail in Sect.~\ref{sec:energyNorm}.

\begin{figure}
\centering
\includegraphics[width=0.9\textwidth]{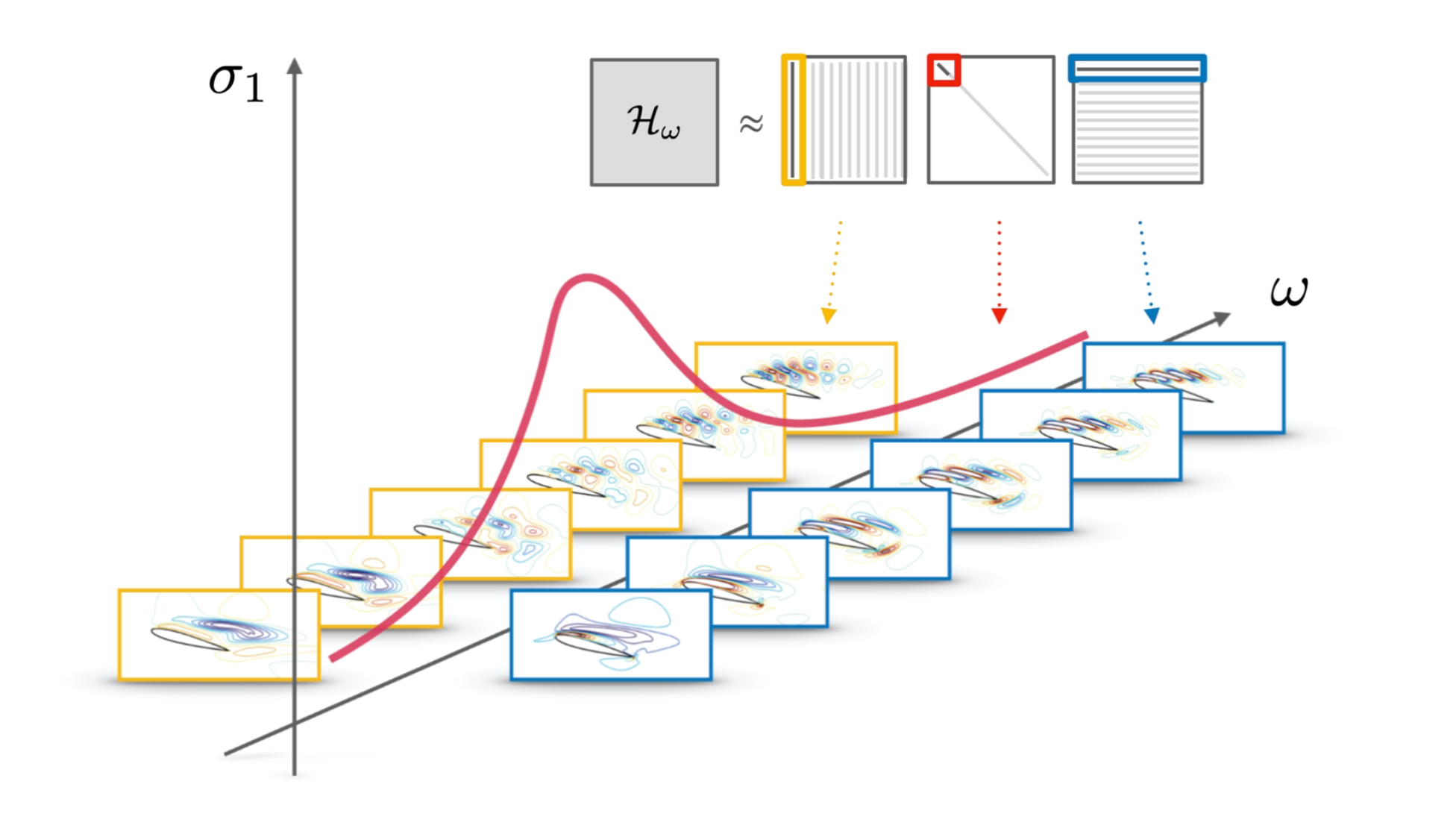}
\caption{Representation of the resolvent analysis.  Blue and yellow boxes show the primary forcing and response modes, respectively.  The amplification/gain distribution is shown in red.}
\label{fig:SVDFFT}
\end{figure}

Equation \ref{eq:optmization} is an optimization problem and the most straightforward way to solve this problem is to perform a singular value decomposition (SVD) of the resolvent operator to find
\begin{equation}
    \mathbf{\mathcal{H}}_\omega=Q_\omega\Sigma_\omega F^*_\omega \mbox{  ,}
    \label{eq_2}
\end{equation}
where the columns of
\begin{equation}
    Q_\omega=[\hat{\mathbf{q}}_{\omega,1}, \hat{\mathbf{q}}_{\omega,2}, ...,\hat{\mathbf{q}}_{\omega,N}] \in \mathbb{C}^{N \times N}
\end{equation}
and 
\begin{equation}    
    F_\omega=[\hat{\mathbf{f}}_{\omega,1}, \hat{\mathbf{f}}_{\omega,2}, ...,\hat{\mathbf{f}}_{\omega,N}] \in \mathbb{C}^{N \times N}
\end{equation}
hold the left and right singular vectors of $\mathbf{\mathcal{H}}_\omega$. Here, the right singular vectors $F_\omega$ represent the optimal basis for the input to the system, that is, the forcing modes. On the other hand, the left singular vectors $Q_\omega$ represent the optimal basis for the output to the system, that is, the response modes. The gains that represent the energy amplification of the system between forcing-response mode pairs are stored on the diagonal of 
\begin{equation}
    \Sigma_\omega=\text{diag}(\sigma_{\omega,1},\sigma_{\omega,2},...,\sigma_{\omega,N}) \in \mathbb{R}^{N\times N} \mbox{  .}
\end{equation}
The formulation of the resolvent analysis is illustrated in Fig. \ref{fig:SVDFFT}.
Superscript $*$ denotes the conjugate transpose of the matrix.  
Matrices $Q_\omega$ and $F_\omega$ are unitary, which means $Q_\omega^{-1} = Q^*_\omega$, 
$Q_\omega Q^* = Q^*_\omega Q_\omega = I$ and  
$F_\omega^{-1} = F^*_\omega$, 
$F_\omega F^*_\omega = F^*_\omega F_\omega = I$.  As the left and right singular vectors are orthonormal, they serve as the basis vectors for the range (output) and domain (input) of the resolvent operator (input-output relation).
Because the forcing and response modes are both normalized (unit length), the information about amplification is entirely contained in the singular values.  Therefore, from Eq. \ref{eq:InputOutput} and given the definition of $Q_\omega$,  $F_\omega$ and  $\Sigma_\omega$, we have:
\begin{equation}
\hat{\mathbf{q}}_\omega=\sum_j\sigma_{\omega,j}\hat{\mathbf{q}}_{\omega,j},\;\;\; \hat{\mathbf{f}}_\omega=\sum_j\hat{\mathbf{f}}_{\omega,j}.
\label{eq:decomp}
\end{equation}

For the singular value decomposition, it is standard to order the singular values in decreasing order such that 
\begin{equation}
    \sigma_{\omega,1} \ge \sigma_{\omega,2} \ge \cdots \ge \sigma_{\omega,N} \ge 0.
\end{equation}
This means that the most amplified mode pair is made by the first forcing mode $\hat{\mathbf{f}}_{\omega,1}$ and first response mode $\hat{\mathbf{q}}_{\omega,1}$ with the amount of amplification (gain) captured by $\sigma_{\omega,1}$.  If the singular values quickly decrease in magnitude, only the first few modes are retained to study the dominant amplification characteristics of the flow. Therefore, the modes that possess relatively large singular values can be kept and those with negligible singular values can be truncated. In fact, many shear-dominated flows have amplification properties that are dominated by the first few resolvent modes.  Thus, we can approximate the resolvent operator as:
\begin{equation}
    \mathbf{\mathcal{H}}_\omega=Q_\omega\Sigma_\omega F^*_\omega
    = \sum_{j=1}^N \hat{q}_{\omega,j} \sigma_{\omega,j} \hat{f}^*_{\omega,j}
    \approx \sum_{j=1}^r \hat{q}_{\omega,j} \sigma_{\omega,j} \hat{f}^*_{\omega,j}
\end{equation}
that can can be considered as a low-rank approximation that retains only the first $r$ modes.  
We can consider that these dominant directions are the most effective or dangerous ways to amplify the perturbations.

The collection of the $j$th gain ($\sigma_{\omega,j}$), response mode ($\hat{\mathbf{q}}_{\omega,j}$), and forcing mode ($\hat{\mathbf{f}}_{\omega,j}$) are referred to as the $j$th resolvent (singular) triplet and is expressed as $(\hat{\mathbf{q}}_{\omega,j},\sigma_{\omega,j},\hat{\mathbf{f}}_{\omega,j})$.  
We remind that the singular value decomposition is performed for each frequency, which means that $\hat{\mathbf{q}}_{\omega,j}$, $\sigma_{\omega,j}$, and $\hat{\mathbf{f}}_{\omega,j}$ are functions of $\omega$ as indicated by their subscripts. However, as these subscripts make the variables notations cumbersome, we drop the subscripts in what follows and refer to them as $\hat{\mathbf{q}}_{j}$, $\sigma_{j}$, and~$\hat{\mathbf{f}}_{j}$.

Generally speaking, what the resolvent analysis seeks is the particular solution to the linear dynamical system, Eq.~{\ref{eq:lin}}, with a harmonic (time-periodic) forcing function.  If there is no forcing applied, then the system is satisfied by the homogeneous solution as the initial value problem.  For an unforced problem, expressing the perturbation variable as
\begin{equation}
    \mathbf{q}^\prime(\mathbf{x,t})
= \hat{\mathbf{q}}(\mathbf{x})e^{\lambda t},
\end{equation}
the linearized governing equation can be transformed to 
\begin{equation}
    \mathcal{L} \hat{\mathbf{q}}(\mathbf{x}) 
    = \lambda \hat{\mathbf{q}}(\mathbf{x}),
    \label{eq:GSA}
\end{equation}
which is an eigenvalue problem for the linearized Navier--Stokes operator $\mathcal{L}$.  Note that the base flow $\overline{\mathbf{q}}$ here needs to be an exact solution to the Navier-Stokes equations in order to have $\mathbf{f}^\prime = \mathbf{0}$.  This is the formulation for the global stability analysis \cite{Theofilis:ARFM11,Theofilis:PAS03}, which provides the eigenvalue $\lambda$ and the eigenvector $\hat{\mathbf{q}}(\mathbf{x})$, known as the (global) instability mode.  As evident from this discussion, the global stability analysis and resolvent analysis complement each other to paint the whole dynamical characterization of the flow.

\subsection{Energy norm}\label{sec:energyNorm}

Since resolvent analysis is concerned with the optimization problem in Eq.~\ref{eq:optmization} that seeks a forcing input that results in the maximum energy amplification and the corresponding flow response, we need to define a physically relevant energy for fluid flow systems. For compressible flows,  both kinetic energy and thermodynamic energy contribute to the total energy in the flow\footnote{For incompressible flows the energy norm can be the kinetic energy
\[
    E=\frac{1}{2}\int_V(u_x{}^\prime{}^2+u_y{}^\prime{}^2+u_z{}^\prime{}^2)dV
\].}, and a commonly used compressible flow energy is that considered by Chu \cite{chu1965energy}.  Chu defined the total disturbance energy in a compressible flow as 
\begin{equation}
    E_\text{Chu}=\frac{1}{2}\int_V\left(\bar{\rho}|\mathbf{u}^\prime|^2+ \frac{\bar{a}^2\rho^\prime{}^2}{\gamma \bar{\rho}}+ \frac{\bar{\rho} c_v T^\prime{}^2}{\bar{T}}\right)dV,
    \label{eq:chu_energy}
\end{equation}
where $\bar{a}=\sqrt{\gamma R \bar{T}}$ is the speed of sound, with $R$ being the gas constant, $\gamma$ the heat capacity ratio and $c_v$ the heat capacity at constant volume.
Recall that variables denoted with $\bar{\cdot}$ indicate the base flow quantities.  

We note that the Chu energy in Eq.~\ref{eq:chu_energy} involves quadratic terms for primitive flow variables, $\mathbf{q}'_p = (\mathbf{u}', \rho', T')$.  To facilitate the use of Chu energy for our state vector $\mathbf{q}'$ in conservative variables, we need a transformation that maps the conservative variables into primitive ones $(\rho^\prime,(\rho u_x)^\prime,(\rho u_y)^\prime,(\rho u_z)^\prime,(\rho e)^\prime)\to(\rho^\prime, u_x^\prime, u_y^\prime, u_z^\prime, T^\prime)$ and vice versa.  That is, 
\begin{equation}
    \mathbf{q}'_p = G\mathbf{q}'
    \quad\text{and}\quad
    \mathbf{q}' = G^{-1}\mathbf{q}'_p,
\end{equation}
where
\begin{equation}
 G= \begin{pmatrix} 
    1 & 0 & 0 & 0 & 0\\ 
     -\dfrac{\bar{u}_x}{\bar{\rho}} & \dfrac{1}{\bar{\rho}}  & 0 & 0 & 0\\
    -\dfrac{\bar{u}_y}{\bar{\rho}} & 0 &  \dfrac{1}{\bar{\rho}}   & 0 & 0\\
   -\dfrac{\bar{u}_z}{\bar{\rho}}   & 0 & 0 &  \dfrac{1}{\bar{\rho}}  & 0\\
     \dfrac{- \bar{p} +\bar{\rho}\bar{U}^2(\gamma-1) }{2R\bar{\rho}^2} &-\dfrac{(\gamma-1)\bar{u}_x}{R\bar{\rho}} & -\dfrac{(\gamma-1)\bar{u}_y}{R\bar{\rho}}  & -\dfrac{(\gamma-1)\bar{u}_z}{R\bar{\rho}}  & \dfrac{(\gamma-1)}{R\bar{\rho}} 
     \end{pmatrix}
\end{equation}
and
\begin{equation}
  G^{-1}= \begin{pmatrix} 
    1 & 0 & 0 & 0 & 0\\ 
    \bar{u}_x & \bar{\rho}  & 0 & 0 & 0\\
    \bar{u}_y & 0 &  \bar{\rho}   & 0 & 0\\
   \bar{u}_z  & 0 & 0 &  \bar{\rho}  & 0\\
     \dfrac{\bar{U}^2}{2}+\dfrac{\bar{p}}{(\gamma-1)\bar{\rho}} & \bar{\rho} \bar{u}_x & \bar{\rho} \bar{u}_y  & \bar{\rho} \bar{u}_z  & \dfrac{\bar{\rho} R}{(\gamma -1)} 
     \end{pmatrix}
\end{equation}
with $\bar{U} = (\bar{u}_x^2+\bar{u}_y^2+\bar{u}_z^2)^{1/2}$.  With the transformation, the Chu energy in Eq.~\ref{eq:chu_energy} can be computed via a numerical integration over the computational domain as
\begin{equation}
    E_\text{Chu}
    =(G\hat{\mathbf{q}})^*W_E(G\hat{\mathbf{q}}),
\end{equation}
where the quadrature weight matrix $W_E$ is given by
\begin{align}
W_E &= \frac{1}{2}
\begin{pmatrix} 
    \Delta V\dfrac{\bar{a}^2}{\gamma \bar{\rho}} & 0 & 0 & 0 & 0\\ 
     0 & \Delta V\bar{\rho} & 0 & 0 & 0\\
     0 & 0 & \Delta V\bar{\rho}  & 0 & 0\\
     0 & 0 & 0  & \Delta V\bar{\rho} & 0\\
     0 & 0 & 0 & 0 & \Delta V\dfrac{\bar{\rho}c_v}{\bar{T}}
     \end{pmatrix},
\end{align}
which contains the cell volume $\Delta V$ and base flow variables at each grid point.  The positive definite weight matrix $W_E$ can be further decomposed through a Cholesky decomposition,  $W_E=F^*_EF_E$, where
\begin{equation}
     F_E  = \sqrt{\frac{1}{2}} 
     \begin{pmatrix} 
   \sqrt{ \Delta V\dfrac{\bar{a}^2}{\gamma \bar{\rho}}} & 0 & 0 & 0 & 0 \\ 
     0 & \sqrt{\Delta V\bar{\rho}}  & 0 & 0 & 0\\
     0 & 0 &\sqrt{ \Delta V\bar{\rho}}  & 0 & 0\\
     0 & 0 & 0 &\sqrt{ \Delta V\bar{\rho} }& 0\\
     0 & 0 & 0 & 0 &\sqrt{ \Delta V\dfrac{\bar{\rho}c_v}{\bar{T}}}
     \end{pmatrix}.
\end{equation}
Finally, this allows us to express the Chu energy as a weighted $L_2$ norm of the conservative state vector, since
\begin{equation}
    E_\text{Chu}
    =(G\hat{\mathbf{q}})^*W_E(G\hat{\mathbf{q}})
    =(F_E G\hat{\mathbf{q}})^*(F_E G\hat{\mathbf{q}})
    = \| F_E G\hat{\mathbf{q}} \|_2^2.
\label{eq:Cholesky}
\end{equation}
This weighted $L_2$ norm for the conservative state vector is referred to as the Chu norm, denoted by
\begin{equation}
    E_\text{Chu} = ||G\hat{\mathbf{q}}||_{E} = \| F_E G\hat{\mathbf{q}} \|_2^2.
\end{equation}

By exploiting the Chu norm, the optimization problem in Eq.~\ref{eq:optmization} that seeks an optimal forcing that maximizes the Chu energy can be converted to a $L_2$-optimization problem, which can be solved via a singular value decomposition of an appropriately weighed resolvent operator \cite{Schmid01}.  To account for both the Chu norm and conservative-to-primitive transformation, Eq.~\ref{eq:InputOutput} becomes
\begin{equation}
    F_E\underbrace{G\hat{\mathbf{q}}_\omega (\mathbf{x})}_{\substack{\text{response} \\\text{mode}}}
    =\underbrace{F_E(-i\omega I-G\mathcal{L}G^{-1})^{-1}F_E^{-1} }_{\text{resolvent operator}}F_E\underbrace{G\hat{\mathbf{f}}_\omega(\mathbf{x})}_{\substack{\text{forcing} \\\text{mode}}}.
\end{equation}
This brings us to the energy-weighted resolvent operator 
\begin{equation}
    \mathcal{H}_{\omega}=F_EG(-i\omega I-\mathcal{L})^{-1}G^{-1}F_E^{-1}.
\end{equation}
Performing the singular value decomposition on this expression of $\mathcal{H}_{\omega}$, the resulting left and right singular vectors are $F_EG\hat{\mathbf{q}}$ and $F_EG\hat{\mathbf{f}}$, which are appropriately weighted for computing the energy norm with an $L_2$-norm and in primitive variables. In order to visualize the actual forcing and response modes over the computational domain, these should be premultiplied by $F_E^{-1}$ also keeping in mind that the solutions are now expressed in primitive variables.

\subsection{Base flow}
\label{sec:baseflow}

The linearization of the Navier-Stokes operator is performed about a chosen base flow ($\mathbf{q}_b$).  Traditionally, the base flow is chosen to be a steady-state solution of the Navier-Stokes equations, which is also referred to as an equilibrium state or a fixed point of the fluid flow system.  In such case, the linearization of Eqs.~\ref{eq:NS-cons} provides Eqs.~\ref{eq:LinearConservative1} to \ref{eq:LinearConservative3}, assuming that the perturbations are small in magnitude.  Steady-state analytical solutions of the Navier-Stokes equations are examples of the equilibrium states, such as the Poiseuille or Couette flow.  

Numerical simulations can also be used to find the equilibrium state as the base flow.  For naturally steady flows, the base flow can be obtained by running the simulations over a sufficiently long time until all unsteadiness subsides to allow for the flow solution to arrive at the stable equilibrium state.  For unstable flows, a time-invariant exact solution to the Navier-Stokes equation can be sought numerically by artificially damping the flow unsteadiness.  The selective frequency damping technique \cite{aakervik2006steady} can be used to numerically damp unsteadiness and determine the unstable equilibrium state.  In this approach, an extra feedback term is added to the right-hand side of the Navier-Stokes equation to suppress any temporal fluctuation, thus forcing the flow towards the steady-state solution.  This is achieved through
\begin{equation}
    \begin{split}
        & \frac{d \mathbf{q}}{d t}=\mathcal{N}(\mathbf{q})-\chi(\mathbf{q}-\mathbf{q}^\star),\\
        & \frac{d \mathbf{q}^\star}{d t}=\frac{\mathbf{q}-\mathbf{q}^\star}{\Delta},
    \end{split}
\end{equation} 
where $\Delta$ is the width of the low-pass filter, $\mathbf{q}^\star$ is the filtered solution and $\chi$ a coefficient that tunes the convergence of $\mathbf{q}$ toward $\mathbf{q}^\star$. Therefore, both $\mathbf{q}$ and $\mathbf{q}^\star$ converge toward the filtered steady solution $\mathbf{q}_{b}$ that satisfies $\mathcal{N}(\mathbf{q}_b)=0$.  This method has been later modified by \citep{jordi2015adaptive} to enable adaptive adjustments of $\Delta $ and $\chi$ whose values are otherwise known a priori.  Another method for computing the unstable equilibrium base flows is a time-stepper formulation of the Newton-GMRES, also called Newton-Krylov methods \citep{dijkstra2014numerical, frantz2023krylov}.  Similar to other gradient-based algorithms, these methods seek the solution $\mathbf{q_b}$ to $\mathcal{N}(\mathbf{q_b})=0$ by iteratively minimizing the residual $\mathbf{r}$ in $\mathcal{N}(\mathbf{q_b^*})= \mathbf{r}$. 

We note that the above numerical approaches to obtaining an unstable equilibrium state generally work for flows of low to modest Reynolds numbers.  However, for high Reynolds number turbulent flows, solving for an unstable equilibrium state can be computationally challenging.  Moreover, there is no guarantee that a steady-state solution exists.  Even if an equilibrium state is successfully sought, it can be quite different from actual turbulent flow or its time-averaged flow field and can limit the interpretability of the results of the resolvent analysis performed about the equilibrium state.  Hence, the use of equilibrium state (fixed point) solutions has been mainly limited to analyses close to the stability thresholds.

\begin{figure}
\centering
\includegraphics[width=0.5\textwidth]{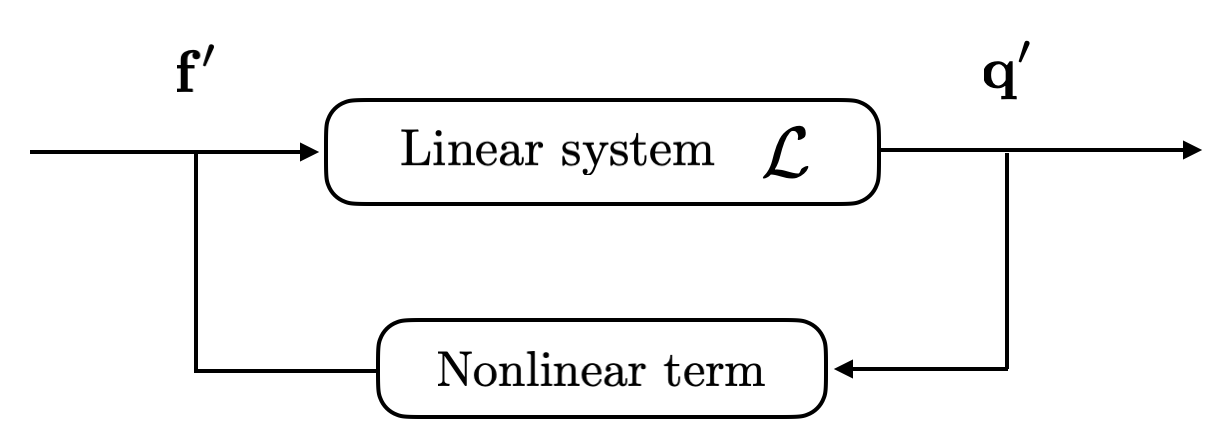}
\caption{\label{fig:feedbackLoop}Feedback loop with the nonlinear term acting as part of forcing.}
\end{figure}

More recently, an alternative perspective for resolvent analysis has emerged to extend its use for base flows beyond an equilibrium solution of the Navier-Stokes equations.  For statistically stationary flows, one can consider the time-averaged flow\footnote{For flows featuring spatial periodicity, it is recommended to perform time- and spatial-averaging along the homogeneous directions.} as the base flow.  In this case, the linearization performed to reach Eqs.~\ref{eq:LinearConservative1} to \ref{eq:LinearConservative3} does not assume the higher-order nonlinear terms to be negligible.  Instead, the nonlinear terms are considered to be part of the forcing term (see Eq.~\ref{eq:forcing}), based on a feedback process as depicted in Fig.~\ref{fig:feedbackLoop} \cite{mckeon2010critical}.  By taking this perspective, the focus is placed on the input-output relationship taking place between the flow state and nonlinear forcing with the resolvent operator as a transfer function.   The use of time-averaged flows in resolvent analysis has offered a tremendous amount of insights into turbulent flow physics \cite{mckeon2010critical, yeh2019resolvent,yeh2020resolvent, sun2020resolvent}.  The time-average flows can be obtained from direct numerical simulations, large-eddy simulations, and Reynolds-Averaged Navier-Stokes simulations.  

When using a base flow (steady-state or time-averaged) whose linear dynamics is unstable, some care is needed in interpreting the results from the resolvent analysis.  In such case, insights on asymptotic input-output amplification characteristics would not be very meaningful since the intrinsic instabilities would become larger than any finite amplification.  However, one can consider examining the input-output amplification characteristics on a time scale shorter than that associated with the dominant instability of $\mathcal{L}$.  This can be achieved by using the exponentially discounting approach to restrict the analysis over a finite time horizon \citep{jovanovic2004modeling,yeh2020resolvent}.   Simple windowing can be applied by multiplying $\exp(-\gamma t)$, where $\gamma > 0$, to both sides of Eq.~\ref{eq:lin}.  It should be noted that the stability of time-averaged flow should be taken with a grain of salt, since stability characterization is not very meaningful for base flows that are not equilibrium states (fixed points).  Nevertheless, the use of discounting provides a conservative reading into the amplification characteristics that emerge over a time scale shorter than that of the modal instability, which instead provides information on the asymptotic amplification characteristics.  Further details on the discounting approach are offered in Sect. \ref{sec:discounted}.  

An alternative perspective is to treat the nonlinear residual term as an eddy viscosity term \cite{morra2019relevance,amaral2021resolvent, pickering2021optimal, symon2023use}.  The eddy viscosity term that models the nonlinear term is informed by the base flow and consequently modifies the linear operator.  Such an approach has been shown to improve the prediction of coherent structures using resolvent modes \citep{morra2019relevance}.  Potentially, the damping effect of the introduced eddy viscosity can also ensure the stability of the overall system.  Another approach to model the nonlinear terms is to seek a linear operator modifier via solving a covariance completion problem \cite{zare2017colour}.  This is also shown to improve the prediction of higher statistical moments using resolvent modes. 

There are also efforts to use experimentally acquired base flow in resolvent analysis \cite{mckeon2010critical}.  Care should be taken to consider the effect of noise in experimental measurements and the possible lack of spatial resolution within the boundary layer.  There are ongoing effort to incorporate data assimilation to resolve these issues \cite{symon2020aiaaj}.

\subsection{Linear operator}
\label{sec:Linop}

Let us describe how the linear operator can be constructed.  Because the size of the linear operator $\mathcal{L}$ is $N \times N$, which is generally very large, special care must be taken to create or gain access to such a matrix. Luckily, $\mathcal{L}$ is generally sparse.  While the matrix can be stored on a personal computer for small-scale problems (1D and some 2D problems), it can be enormous for larger-scale 2D (biglobal) and 3D (triglobal) problems requiring access to high-performance computing resources.  This is especially the case for higher Reynolds numbers and external flows that require a large number of grid points to capture the flow physics.

Once the linear operator $\mathcal{L}$ is in hand, the resolvent operator is available through $\mathcal{H}_\omega = [-i\omega I - \mathcal{L}]^{-1}$.  However, we stress that {\it{the matrix inverse should not be explicitly computed}}.  Numerical algorithms to find a matrix inverse suffer from numerical instabilities and gross inaccuracies, besides being prohibitive regarding the computational cost and memory requirement. Later we discuss how to find the dominant resolvent modes and gains without relying on the availability of the matrix inverse.

There are mainly two approaches to establishing the linear operator within a computer program.  The first approach is to develop a computer program that returns the action of the discrete linear Navier-Stokes operator $\mathcal{L}(\cdot)$ on a given vector.  Such a program is equivalent to a right-hand-side function of the linearized Navier-Stokes equations \ref{eq:linear_jacobian}.
In Appendix \ref{AppendixA}, we provide full details of the linearized compressible Navier-Stokes equations for the three-dimensional (triglobal) case.  Simplifications can be achieved for two-dimensional (biglobal) base flows, in which case the homogeneous direction can be expressed in terms of a spatial Fourier series.  Details on the biglobal case are also offered in the appendix. This approach avoids explicitly forming a matrix for the linear operator and can free up the computer memory and storage space.  However, to facilitate such a matrix-free approach, the adjoint linearized Navier-Stokes operator/program is also needed for the computation of the singular value decomposition of the resolvent operator.  More details on such a matrix-free approach and the need for the adjoint operator will be discussed in Sect.~\ref{sec:Methods}.

The other approach is to explicitly form the matrix, and is called the matrix-forming method. 
However, deriving the matrix for the discrete version of the linearized Navier-Stokes equations can be a major task by itself, especially for complex grid structures and for biglobal and triglobal flows.  There are two different ways to explicitly derive the linear operator matrix which are described below.  We stress that explicitly computing the linear operator matrix, which implies storing the matrix, can be expensive in terms of memory and storage allocation, especially for large-scale problems since the matrix dimension scales as $O(N_\text{cell}^2)$ while being sparse.  However, having access to the actual matrix can facilitate the access also to the adjoint calculations.  This can be helpful when the treatment of the boundary conditions is non-trivial for the adjoint operator.

\begin{figure}
\centering
\includegraphics[width=\textwidth]{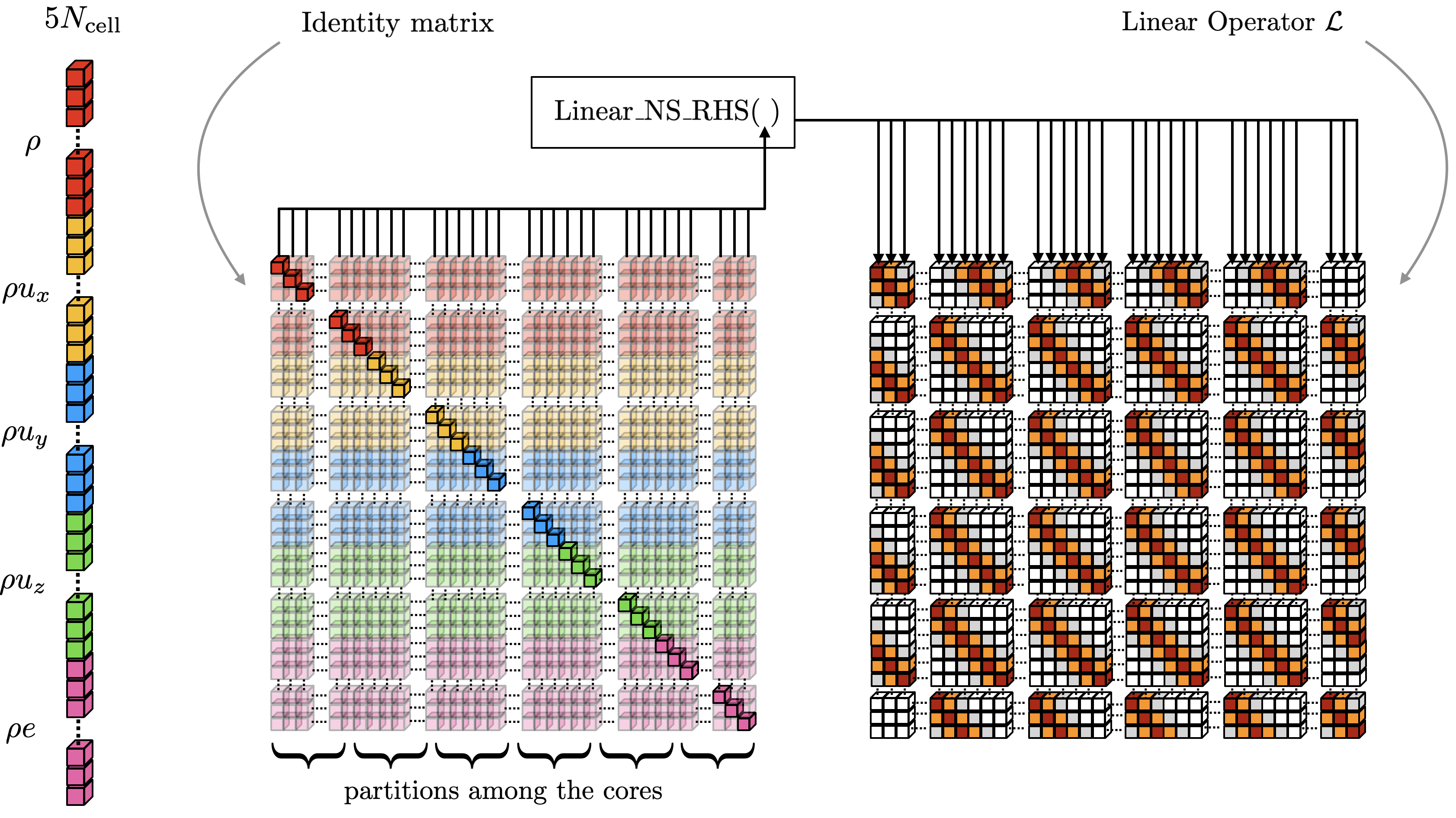}
\caption{\label{fig:LieqOp}
Extraction of the linear operator $\mathcal{L}$ from a linearized NS code.}
\end{figure}

\paragraph{Matrix extraction}
If a computer code \texttt{Linear\_NS\_RHS($\cdot$)} is available to evaluate the right-hand side of the linearized NS equations based on Eqs.~\ref{eq:LinearConservative1} to \ref{eq:linPrimitivevariable}, we can form a matrix for the linearized Navier-Stokes operator $\mathcal{L}$ from such a code.  This process can be viewed as right-multiplying an identity matrix $I$ to $\mathcal{L}$.  In practice, we can do so by passing one column of $I$ at a time to the linearized Navier-Stokes operator.  The passing of the $i$th column of the identity matrix to the linear operator (\texttt{Linear\_NS\_RHS}($I_i$)) returns the corresponding $i$th column of the linear operator $\mathcal{L}$. This process gives the values of the linear operator in each position as illustrated in Fig.~\ref{fig:LieqOp}, where the action of the linear operator on the identity matrix $I$ is depicted.  While this may appears as a computationally cumbersome process, the parallelization of such a task allows for high level of scalability by taking advantage of the sparsity of $\mathcal{L}$ and the the non-overlapping stencil within the operator. 

\paragraph{Fr\'echet derivative}
An alternative way to explicitly compute the linear operator is by using a classic nonlinear CFD solver.  This is useful when a solver is not available for the linearized NS equations. In this case, the linear operator $\mathcal{L}$ can be found through a Fr\'echet derivative of the nonlinear operator $\mathcal{N}$ through
\begin{equation}
    \mathcal{L}I_i=\frac{\mathcal{N}(\mathbf{q}_b+\epsilon I_i)-\mathcal{N}(\mathbf{q}_b-\epsilon I_i)}{2\epsilon},
\end{equation}
where $\epsilon\ll 1$ \citep{tezuka2006three, mack2008global, an2011finite, de2012efficient} and $I_i$ the i-th column vector of the identity matrix. This formulation is based on the second-order finite difference, but any order can be used.

\subsubsection{Remarks on computational setup}
A comment should be made on the boundary conditions, which are problem-dependent. Nonetheless, it is essential to consider that for most cases, perturbations $q'$ are expected to vanish in the far field, making the use of Dirichlet boundary conditions an appropriate option. Dirichlet boundary conditions are also used on the body walls, while a Neumann boundary condition should be used at the outflow.
These boundary conditions are generally embedded as part of the linear operator (discretization scheme).  
For external flows, the use of sponge zones at the exterior (e.g., inlet, outlet and far-field boundaries) of the computational domain is also a valuable option to avoid possible numerical problems given by the transpose (adjoint) operator.  We note that the treatment of boundary conditions for incompressible external flows can be challenging compared to those of compressible flows.  This is due to the mass conservation acting as a constraint to the overall system.  Such an issue may not appear for settings such as internal flows.

Another matter to consider is the spatial grid to perform the resolvent analysis.  The computational grid and domain size used for the resolvent analysis do not necessarily need to match that of those used to determine the base flow.   For base flows determined from direct numerical simulations or large eddy simulations, the grid resolutions should be very fine where small-scale flow structures appear.  If the resolvent analysis is used to extract only the dominant modes, their structures generally are larger than the small turbulent eddies, which allows us to use a grid coarser than the one used for obtaining the base flow.

For external flows, the domain size for resolvent analysis can be much smaller in size as long as the primary forcing and response modes fit well within the smaller domain.  It should be noted that the forcing modes tend to emerge upstream of the response modes, which can require grid refinement to be performed in the upstream region (similar to adjoint simulations).  Computational savings can be attained through grid coarsening and the usage of smaller domains for resolvent analysis.  However, such computational grid/domain setups should be carefully chosen and verified to ensure that the discretizations do not influence the final outcome.  Further discussion on these matters can be found in Sect. \ref{sec:set-up}.

\begin{figure}
\centering
\includegraphics[page=1,trim=0mm 0mm 0mm 0mm, clip,width=0.85\textwidth]{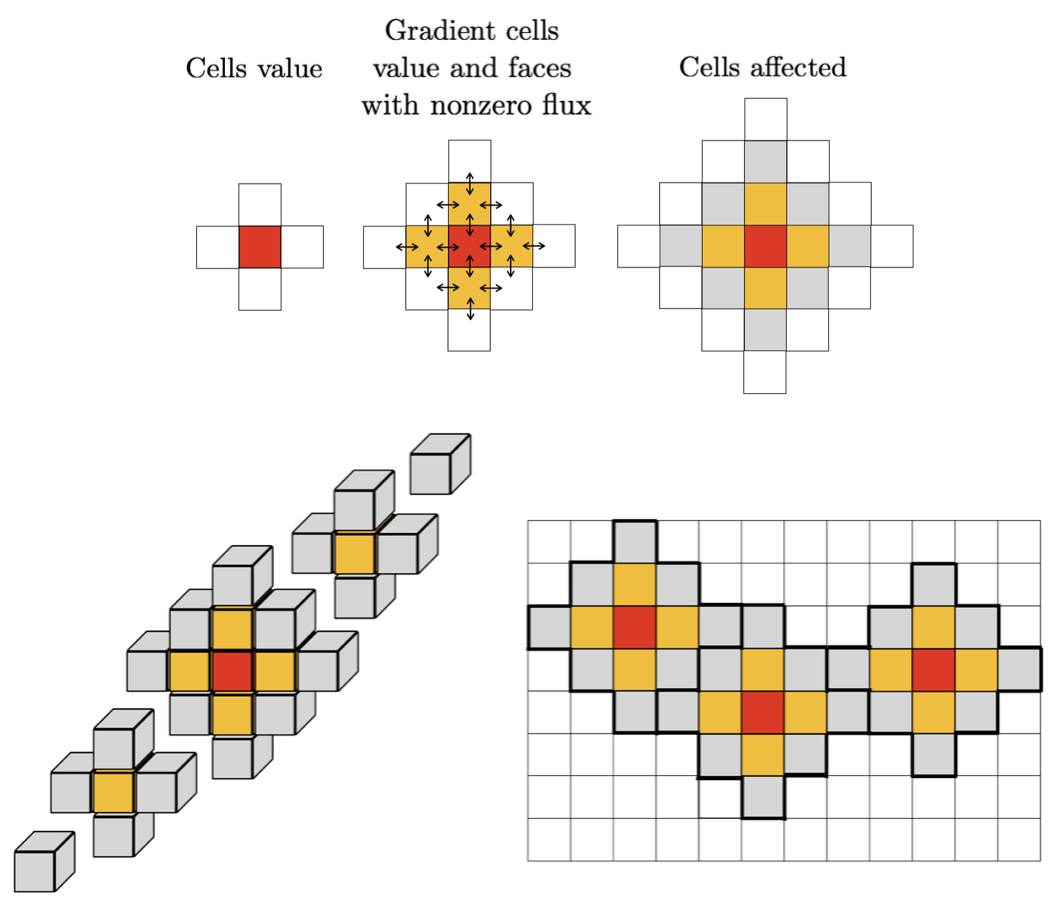}
\put(-325,270){(a)}
\put(-325,130){(b)}
\put(-160,130){(c)}
\caption{\label{fig:Stencil}(a) Stencil of the second-order finite volume formulation. Blank cells in the stencil are 0-valued cells. (b) three-dimensional and (c) two-dimensional stencils with representation of reduced cost and matrix compression.}
\end{figure}

If the computation of the linear operator $\mathcal{L}$ is expensive, parallelization can be used to accelerate the calculation.
A useful parallelization when explicitly computing the linear operator matrix can be achieved by considering that each evaluation (${\tt Linear\_NS\_RHS}(I_i)$ in Fig.~\ref{sec:Linop}) is independent from the others.
Moreover, in case of a structured mesh, another way for decreasing the computational cost takes into account the structural orthogonality of the linear operator matrix. This orthogonality corresponds to columns of the linear operator matrix that do not have a nonzero in a common row \cite{gebremedhin2005color,mettot2013linear}, which depends on the chosen numerical scheme. By identifying the structurally orthogonal columns of the linear operator, concurrent evaluations such as ${\tt Linear\_NS\_RHS} (I_i+I_j+I_k)$ in Fig.~\ref{fig:LieqOp} can be assessed. This means simultaneously perturbing cells that are sufficiently distant from each other to avoid mutual influence. 
The nonzero elements in each column of the linear operator matrix depend on the stencil of the operator (Fig. \ref{fig:Stencil}.(a)), which is the ensemble of cells that are affected by a variation in the central cell, shown in red. The red cell represents the perturbation of the $m$th variable in the $l$th cell, correspondent to the $i$th column of the identity matrix, $I_i$, where $i=l+(m-1)N_{\text{cell}}$.
By determining the stencil from the numerical scheme (Fig. \ref{fig:Stencil}.(b)), we are able to identify cells that do not affect each other when perturbed (yellow and gray cells do not intersect) as depicted in Fig. \ref{fig:Stencil}.(c), where the evaluation of ${\tt Linear\_NS\_RHS}(I_i+I_j+I_k)$ is illustrated.  


\subsection{Extensions of resolvent analysis}\label{sec:Variations}
Over the past few years, there have been a number of extensions made to the resolvent analysis. Here, we present notable extensions that can be useful in focusing the analysis on certain flow physics and addressing some challenges associated with the basic resolvent analysis formulation.

\subsubsection{Spatially windowed resolvent analysis}\label{sec:Window}

In some cases, we are interested in understanding the input-output relationship over a specific region of the flow.  This can be achieved by introducing spatial windows to the resolvent analysis formulation \cite{juenPOF2016,kojima2020resolvent,yeh2020resolvent}.  Here, we incorporate a spatial window $B$ to the forcing input ${\mathbf{f}}'$ and an examination window $C$ to the flow response ${\mathbf{q}}'$ to yield an output ${\mathbf{y}}$.  The resulting set of equations become 
\begin{equation}\label{eq:SystWindowed}
    \begin{split}
    \frac{d \mathbf{q}'}{d t} &=\mathcal{L}\mathbf{q}' + B\mathbf{f}',\\
    \mathbf{y} &= C\mathbf{q}',
    \end{split}
\end{equation}
where matrices $B$ and $C$ are diagonal matrices whose entries are $1$ if the cell resides in the window and $0$ otherwise. This system is the so-called linear state-space representation in dynamical systems theory, which can serve as the basis to invite the use of extensive theories and algorithms from the field of controls and dynamical systems \cite{MurrayAstrom, Friedland, DullerudPaganini}.   
Note that system \ref{eq:SystWindowed} reduces to Eq. \ref{eq:lin} if $B=C=I$ which means the whole spatial domain is the area of interest for both the forcing and response.

Following the same procedure in Sect. \ref{sec:inputoutput}, we can derive the transfer function between the force input $\hat{\mathbf{f}}$ and output $\hat{\mathbf{y}}$ at frequency $\omega$ as
\begin{equation}
    \mathcal{H}_{\omega}=C[-i\omega I-\mathcal{L}]^{-1}B,
\end{equation}
which, using the compressible NS equations in conservative form, takes the following form:
\begin{equation}
    \mathcal{H}_{p,\omega}=F_ECG[-i\omega I-\mathcal{L}]^{-1}G^{-1}BF_E^{-1}
\end{equation}
by taking into account the energy norm and the transformation from the conservative to primitive variables.

The windowed resolvent analysis enables not only the examination of flow physics in certain regions but also those that may be buried under the dominant flow dynamics.  Kojima et al.~\cite{kojima2020resolvent} considered the use of windowing to examine the possible emergence of buffet for transonic airfoil flows in low-Reynolds number flight conditions.  Although the examination of the entire flow field will reveal the wake unsteadiness to be dominant, windowing the analysis to the region around the normal shock over the wing uncovers the input-output dynamics for buffeting.

Spatial windowing is a valuable tool for exploring the input-output relationships between particular spatial regions of interest. As an example, windowing has been used to study the aeroacoustics of high-speed isothermal turbulent jets \cite{Jeun:PF16input}. In this analysis, windowing was applied through matrix $B$ to confine the inputs to a region near the jet nozzle, while the matrix $C$ was applied to the farfield to restrict the output to the isentropic region where acoustic waves propagate. This analysis unveiled the source mechanisms that contribute to jet noise generation. When employing spatially windowed resolvent analysis, the linear operator usually covers a larger domain, but the analysis focuses on a smaller area. 

Another use of windowed resolvent analysis is the componentwise resolvent analysis \citep{jovanovic2005componentwise,schmid2007nonmodal}, which examines the maximization of the energy through specific variables of the forcing and response modes.  For example, one can study the maximization of energy contribution given either the cross-stream or streamwise velocity component.  In this case, matrices $B$ and $C$ would be matrices whose diagonal entries are $1$ for the components of the state vector that we aim to maximize the energy with, while entries are $0$ for the other variables.

\subsubsection{Temporally windowed (discounted) resolvent analysis}\label{sec:discounted}

The original resolvent analysis reveals the input--output relationship that is meaningful only if the system itself is stable.  If the system of interest is unstable, the state eventually diverges, even without the action of the forcing function.  The system response to the forcing input can become buried under the exponentially growing instability.  Although such a situation may discourage the use of resolvent analysis, it can still provide great insights into the system dynamics with only a simple extension.

To address the case for which we have an unstable system, consider filtering the forcing and response modes by $h = \exp(-\gamma t)$, where $\gamma > 0$.  If the parameter $\gamma$ is chosen such that $\gamma > \max \text{Real}[\lambda(\mathcal{L})]$, the largest real component of the eigenvalues of $\mathcal{L}$, the filtered response, 
\begin{equation}
    \check{\mathbf{q}} \equiv h \hat{\mathbf{q}} = \exp(-\gamma t) \hat{\mathbf{q}},
\end{equation} 
remains bounded over time.  The parameter $\gamma$ is referred to as the (temporal) discounting parameter \cite{jovanovic2004modeling} and it enables us to examine the system dynamics on a finite time scale shorter than that associated with the dominant instability.  The filter function $h$ acts here as a temporal window.

This discounting procedure modifies the input-output relationship in Eq.~\ref{eq:InputOutput} to become
\begin{equation}
\label{eq:InputOutput_discounted}
    \check{\mathbf{q}}(\mathbf{x})
    =\mathbf{\mathcal{H}}_{\omega,\gamma} \check{\mathbf{f}},
\end{equation}
where
\begin{equation}
    \mathbf{\mathcal{H}}_{\omega,\gamma} \equiv [(-i\omega + \gamma) I -\mathcal{L}]^{-1}
\end{equation}
is the discounted resolvent operator.  Given this discounted resolvent formulation, the decomposition of $\mathbf{q}'$ and $\mathbf{f}'$ is not performed by using the Fourier transform (suitable for asymptotic input-output analysis).  What we have now is the Laplace transform that relates the discounted modes in the frequency domain to the temporal domain \citep{jovanovic2004modeling,jovanovic2005componentwise}.

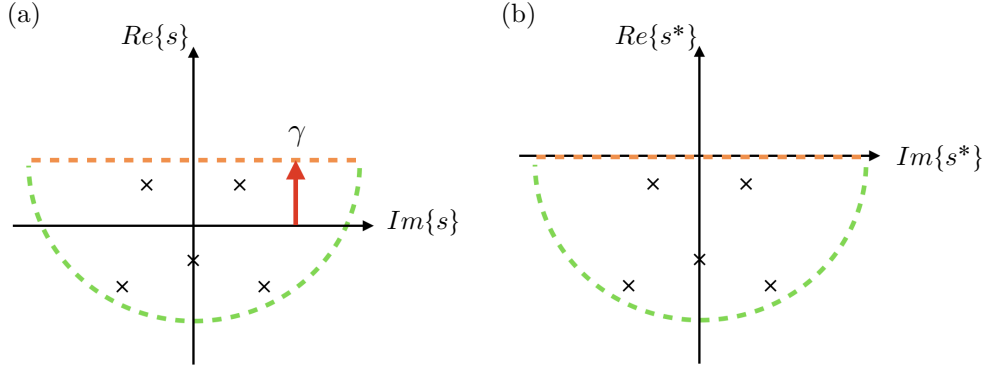
\begin{figure}[t]
\centering
\begin{tikzpicture}
    
\node[anchor=south west,inner sep=0] (image) at (0,0) {\includegraphics[width=0.9\textwidth]{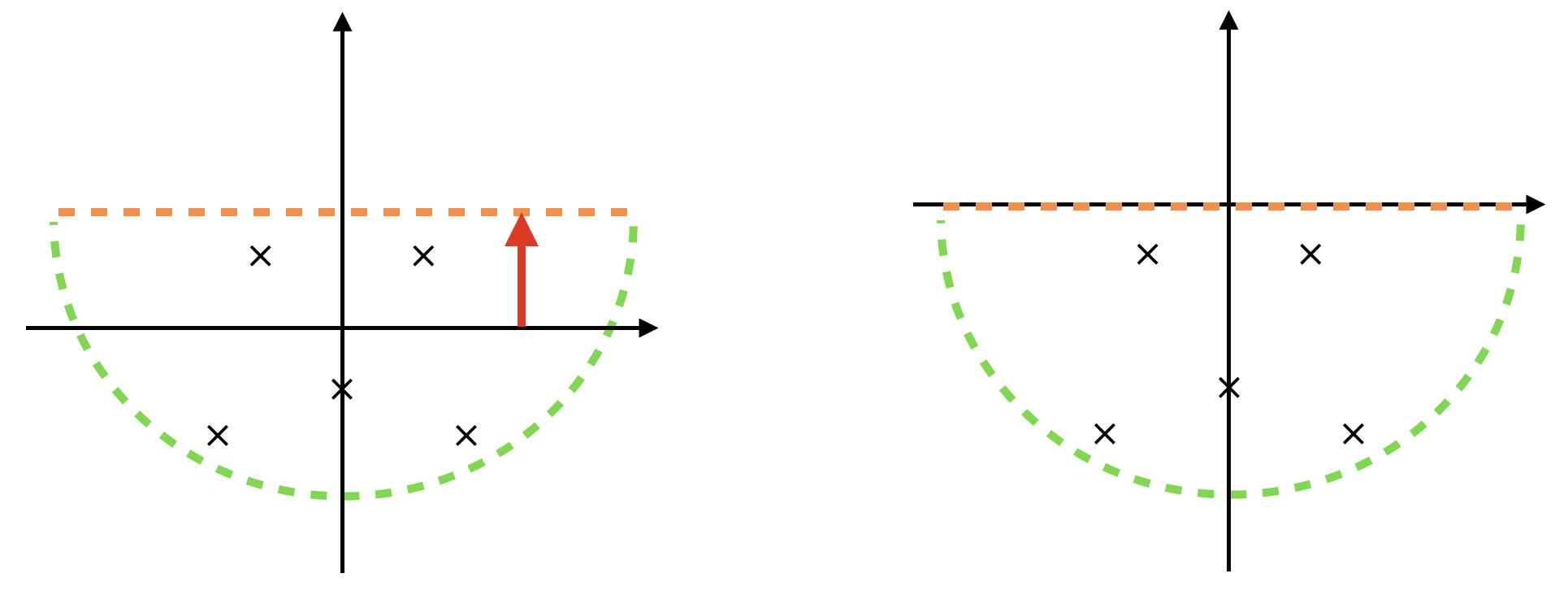}};
\node[anchor=west] at (0.0,4.8) {(a)};
\node[anchor=west] at (6.5,4.8) {(b)};
\node[anchor=west] at (3.7,3.2) {{\color{black}{\large{$\gamma$}}}};

\node[anchor=west] at (1.5,4.5) {{\color{black}{$Re\{s\}$}}};
\node[anchor=west] at (5,2.05){{\color{black}{$Im\{s\}$}}};
\node[anchor=west] at (8,4.5) {{\color{black}{$Re\{s^*\}$}}};
\node[anchor=west] at (11.7,2.9){{\color{black}{$Im\{s^*\}$}}};

\end{tikzpicture}
\caption{\label{fig:DiscountedInt}Effect of discounting for (a) $\mathcal{H}_{s}$ and (b) $\mathcal{H}_{s^*}$. The discounted resolvent analysis for unstable $\mathcal{L}$ can be related to the shift of the vertical integral path in the inverse Laplace transform. Crosses represent poles, \textit{i.e.} the eigenvalue of the linear operator, the dashed orange line represents the integration path while the dashed green line is the closing integration path forming the Bromwich~contour. }
\end{figure}

\begin{figure}
\centering
		\begin{overpic}[scale=0.53]{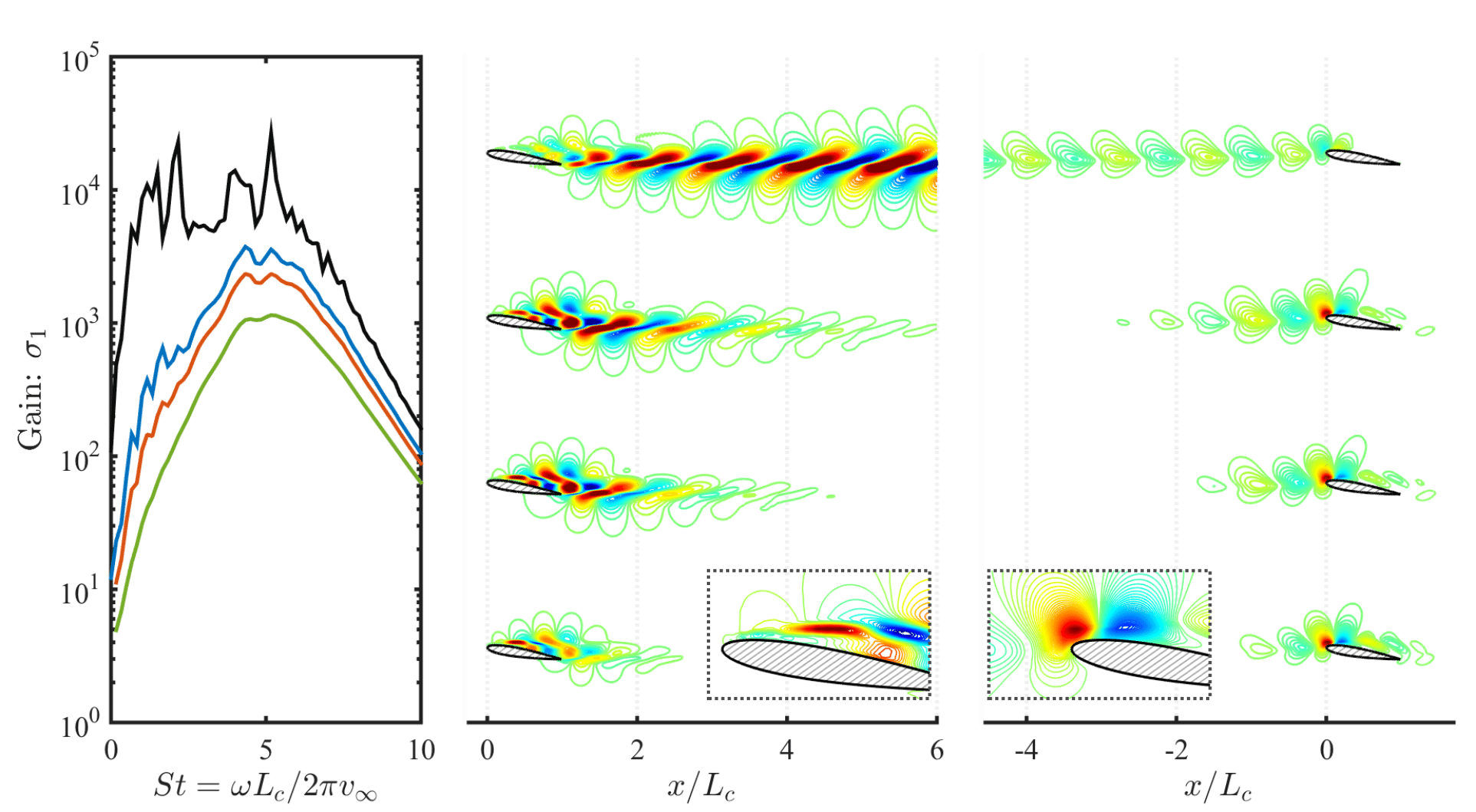}
			\put(1, 53){\small (a)}
			\put(30, 53){\small (b) Response modes}
			\put(65, 53){\small (c) Forcing modes}
			\put(13, 15){\small
							\begin{tabular}{r l} 
								\multicolumn{2}{c}{$\gamma L_c/U_\infty$} \\ \hline\vspace{-0.10in}\\
								{\color{black}--}&{\color{black}$0$} \\
								{\color{blue}--}&{\color{blue}$0.143$} \\
								{\color{red}--}&{\color{red}$0.2$} \\ 
								{\color{green}--}&{\color{green}$0.333$}
							\end{tabular}}
			\put(31, 48.4){\small$\gamma L_c/U_\infty = 0$}
			\put(31, 37.1){\small\color{blue} $\gamma L_c/U_\infty = 0.143$}
			\put(31, 25.8){\small\color{red} $\gamma L_c/U_\infty = 0.2$}
			\put(31, 14.5){\small\color{green} $\gamma L_c/U_\infty = 0.333$}

			\put(82.5, 48.4){\small$\gamma L_c/U_\infty = 0$}
			\put(82.5, 37.1){\small\color{blue} $\gamma L_c/U_\infty = 0.143$}
			\put(82.5, 25.8){\small\color{red} $\gamma L_c/U_\infty = 0.2$}
			\put(82.5, 14.5){\small\color{green} $\gamma L_c/U_\infty = 0.333$}
   
		\end{overpic}
	\caption{\label{fig:yeh_jfm2019_discount} Temporal-windowed resolvent analysis of a separated flow over an airfoil with different choices of $\gamma$. (a) Gain distribution over frequency in $St$; the leading resolvent (b) response and (c) forcing modes at $St = 0.833$.  The streamwise extent of the modal structures shortens with increasing $\gamma$.}
\end{figure}
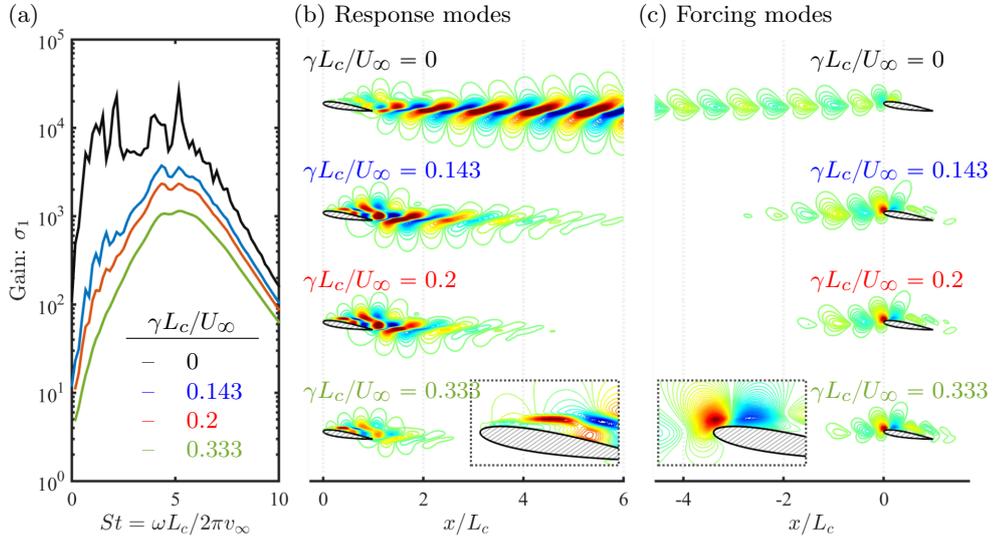

Using the Laplace representation we have
\begin{equation}
    \begin{split}
    \mathbf{q}'(\mathbf{x,t})=\frac{1}{2i\pi}\int^{\gamma+i\infty}_{\gamma-i\infty}\hat{\mathbf{q}}_s(\mathbf{x})e^{st}ds
\end{split}\label{eq:laplace}
\end{equation}
instead of Eq.~\ref{eq:FourierTrans}, where $\gamma$ is greater than the real parts of all the possible singularities of $\hat{\mathbf{q}}_s=\mathcal{H}_s\hat{\mathbf{f}_s}$, where  
\begin{equation}
\mathcal{H}_s=[s I-\mathcal{L}]^{-1} .
\end{equation}
Therefore, the singularities of the system coincide with the poles of the transfer function, which are the eigenvalues of the linear operator.   The integration along a line above the singularities (orange line in Fig. \ref{fig:DiscountedInt}.a) ensures convergence of the integral over a semi-indefinite time domain, $t\in [0,\infty)$.  With a change of variable $s^*=s-\gamma$, we recover the Fourier transform
\begin{equation}
    \begin{split}
    \mathbf{q}(\mathbf{x,t})=e^{\gamma t}\frac{1}{2i\pi}\int^{i\infty}_{-i\infty}\hat{\mathbf{q}}_{s^*}(\mathbf{x})e^{s^*t}ds^*,
\end{split}
\end{equation}
where 
\begin{equation}
\hat{\mathbf{q}}_{s^*}=\mathcal{H}_{s^*}\hat{\mathbf{f}}_{s^*} 
\quad
\text{and}
\quad
\mathcal{H}_{s^*}=[(s^*+\gamma) I-\mathcal{L}]^{-1} .
\end{equation}
The discounted transfer function $\mathcal{H}_{s^*}$ has now only stable poles and corresponds to the system in Fig.~\ref{fig:DiscountedInt}.b.  Essentially, discounting acts as a shift in the complex plane, affecting only $\mathcal{H}$, while not damping unstable eigenvalues in $\mathcal{L}$ spectrum.  Increasing the value of time-window constant, $\gamma$, translates to a shorter time scale over which the input-output dynamics is examined.  The shorter time window for the perturbation to grow in amplitude will result in lower resolvent gain.  For external flows dominated by advective physics, the increase in $\gamma$ can also result in forcing and response modes of shorter streamwise extent.  These effects of the time window are demonstrated in Fig.~\ref{fig:yeh_jfm2019_discount} \cite{yeh2019resolvent}, where we observe that the increase in $\gamma$ results in a decrease in gain and spatially shortened modal structures in the streamwise direction. 

In the above discounted resolvent analysis, we observed that temporal windowing through an exponential function led to the replacement of the Fourier transform with the Laplace transform to define the forcing and response modes. This can be further generalized to accommodate other types of temporal windowing.  In recent studies \cite{BallouzSciTech2023, BallouzArxiv2023,lopez2023sparsity}, a wavelet transform was used to temporally window the input-output relationship.  This analysis has enabled the use of resolvent analysis to be performed not only on time-invariant base flows but also to time-varying base flows.

\subsubsection{Sparse resolvent analysis}

The forcing modes uncovered from the aforementioned resolvent analysis possess spatially global flow structures.  This means that the forcing modes exhibit spatial oscillations over some regions of the flow.  When the resolvent analysis is used in the hope of guiding flow control efforts, identifying the most appropriate location for an actuator can be difficult.  This is because the standard resolvent analysis is formulated on the use of an $L_2$ norm (see Eqs.~\ref{eq:optmization} and \ref{eq:Cholesky}) which determines a spatially global distribution of the modes.  

\begin{figure}
\centering
\includegraphics[width=\textwidth]{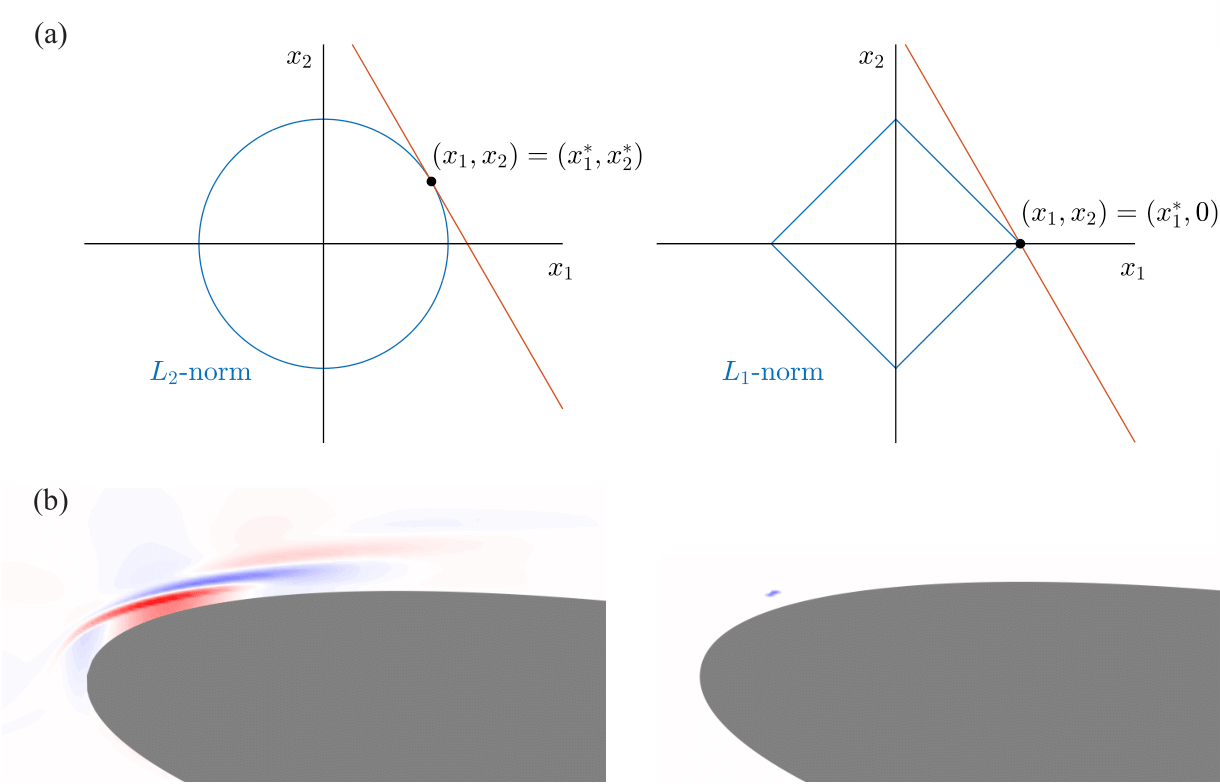}
\caption{(a) Representations of the $L_2$ and $L_1$-norm based optimizations, where the intersection points represent the solutions.  A sparse solution is obtained by the $L_1$-based approach for this two-dimensional example. (b) Sparsification of the forcing mode for an example of flow over an airfoil \cite{skene2022sparsifying}.  Notice the global forcing mode (left) collapsing to a single spatial point for sparse resolvent analysis (right).}
\label{fig:sparse}
\end{figure}

This issue can be addressed by revisiting the use of the $L_2$ norm and instead utilizing a sparsity-promoting norm, such as the $L_1$ norm, in the resolvent analysis framework (see Fig. \ref{fig:sparse}(a) for the general concept).  The $L_1$ norm-based formulation can be used to spatially sparsify the forcing modes.  By doing so, the forcing modes can reveal the exact position where a spatially sparse forcing input can provide the largest amplification in the response mode \cite{skene2022sparsifying}.  Practically speaking, such forcing location amounts to a single point in space. However, it should be noted that a sparse response mode should not be sought.  It would not be physically appropriate to constrain the response mode to be sparse in space.

To implement the sparse resolvent analysis, we can reformulate the optimization problem of Eq.~\ref{eq:optmization} with an alternate setup that seeks to maximize the gain for the $L_1$ norm.  The solution to this sparse resolvent problem would not involve an SVD but can be found with a Riemannian optimization technique.  One caveat to this optimization solver is the need to regularize the $L_1$ norm in the vicinity of the optimal solution to smooth out its gradient, which can be achieved by using a Huber norm \cite{skene2022sparsifying}.

It is interesting that the use of $L_1$ norm not only finds the spatially sparse forcing mode but also uncovers the sparsity in the state variables.  For example, if there are five variables in the state variables as in the case of compressible flow, sparse resolvent analysis can identify which variable out of the state variables are important for forcing.  This is also an important piece of insight in active flow control because it will advise on how the forcing input should be introduced for optimal effects – e.g., either a momentum-based or thermal-based control input.

The use of the $L_1$ norm by itself does not guarantee that the forcing mode will find a spatial point in space that is physically possible.  For instance, the forcing mode from the sparsity-promoting resolvent analysis may suggest actuation to be introduced in the middle of the flow.  In the work of Skene et al.~\cite{skene2022sparsifying}, they performed sparse resolvent analysis for flow over a NACA0012 airfoil, as shown in Fig. \ref{fig:sparse}(b).  Their analysis pinpointed the actuation location and revealed the best combination of forcing variables (momentum injection).  They also revealed the optimal forcing direction for such a forcing input.  Furthermore, they were able to find the optimal forcing point on the airfoil surface by restricting the forcing mode to reside only on the surface (see windowed resolvent analysis).  

\subsubsection{Frequency cross-talks}

In this paper, we limit our discussion to the resolvent analysis that focuses on the input-output dynamics only at a single frequency.  This could restrain the applicability of resolvent analysis for some turbulent flows, where the nonlinear interactions across multiple frequencies can lead to significantly different perturbation growth from that predicted by single-frequency resolvent analyses \cite{brunton2015closed}.  Motivated by these issues, recent efforts have enabled resolvent analysis to account for the interactions across multiple spatial and temporal frequencies at a linear limit by considering base flows with spatial \cite{Chavarin:AIAAJ2020} and temporal periodicity \cite{Padovan:JFM2020harmonic}.
A weakly nonlinear \citep{dwivedi2022oblique} and fully nonlinear \citep{rigas2021nonlinear} resolvent formulations have also been introduced to analyze higher-order perturbation dynamics in the frequency domain.

\subsection{Parametric sensitivity of the resolvent gain}
\label{sec:Gradient}

Performing resolvent analysis for a sweep of parameters can be somewhat computationally taxing.  In this case, identifying the sensitivity (gradients) of the resolvent gain can complement the analysis and enable spline interpolation to provide an accurate mapping of the gain distribution over the parameter space. A sensitivity-based technique \cite{de2014parametric, skene2019adjoint} can be used to approximate variations in the gain $\sigma$ with respect to changes in any parameter.  This approach can significantly reduce the number of singular value decompositions to be performed, because a gradient-informed interpolation can be used. 

This formulation was originally derived for direct and adjoint global stability modes. Because the resolvent forcing modes are similar in spirit to adjoint modes, such as response modes resembling characteristics of the direct global stability modes, this approach can be extended to compute gradients within the resolvent analysis framework. For resolvent analysis, the solutions of both the direct and adjoint problem are computed, with the latter being strictly related to the sensitivity/gradient of the system to variations over the parameters. For this reason, the inner product of the response and forcing modes that are appropriately weighted (energy norm) gives us the gradient/sensitivity with respect to the different parameters. 

The resolvent gain is affected by subtle changes in the temporal frequency, the wavelength, and the base flow characteristics. Specifically, for the gradients of  $\sigma_j$ with respect to frequency and wavelength, the parametric sensitivities can be identified through the forcing and response mode characteristics.

The parametric sensitivity formulation of the $j$th singular value is provided with following approximation \citep{de2014parametric}
\begin{equation}
    \delta\sigma_j
    \approx 
    \text{Real}\left\{\hat{\mathbf{q}}_j^*\delta \mathcal{H} \hat{\mathbf{f}}_j\right\}
\end{equation}
which can also be expressed in terms of the change in the inverse of the resolvent operator $\delta\mathcal{H}^{-1}=\delta[-i\omega I-\mathcal{L}]$ rather than $\delta\mathcal{H}$ such that 
\begin{equation}
    -\frac{\delta\sigma_j}{\sigma_j^2}
    \approx 
    \text{Real}\left\{\hat{\mathbf{f}}_j^*\delta[ -i\omega I-\mathcal{L}] \hat{\mathbf{q}}_j\right\}.
\end{equation}
From this expression, we find that the gradient of $\sigma$ with respect to a parameter $\alpha$ is given by
\begin{equation}
    \frac{1}{\sigma_j^2}\frac{\partial \sigma_j}{\partial \alpha}
    \approx 
    \text{Real}\left\{\left<\hat{\mathbf{f}}_j,\frac{\partial(i\omega I+\mathcal{L})}{\partial \alpha}\hat{\mathbf{q}}_j\right>_E\right\}.
\end{equation}
The above equation now allows us to find the gradient of the resolvent gain with respect to parameters in the resolvent operator.

For the gradient with respect to the forcing frequency $\omega$, the above gradient expression simply becomes
\begin{equation}
    \frac{1}{\sigma_j^2} \frac{\partial \sigma_j}{\partial \omega}
    \approx 
    -\text{Imag}\left\{ \left< \hat{\mathbf{f}}_j,
    \hat{\mathbf{q}}_j \right>_E\right\}
\end{equation}
while the real part would correspond to the gradients with respect to the discount parameter $\gamma$ 
\begin{equation}
    \frac{1}{\sigma_j^2} \frac{\partial \sigma_j}{\partial \gamma}
    \approx 
    \text{Real}\left\{ \left< \hat{\mathbf{f}}_j,
    \hat{\mathbf{q}}_j \right>_E\right\}.
\end{equation}

For other parameters, the parametric sensitivity reduces to 
\begin{equation}
     \frac{1}{\sigma_j^2}\frac{\partial \sigma_j}{\partial \alpha}
     \approx \text{Real}\left\{\left<\hat{\mathbf{f}}_j,
     \frac{\partial\mathcal{L}}{\partial \alpha}\hat{\mathbf{q}}_j\right>_E\right\}
\end{equation}
because $\omega$ is independent of other parameters.  
For example, parametric sensitivity can also be determined for wavenumbers in periodic directions of the problem.  For a biglobal setup (see Appendix \ref{AppendixA}), the parametric sensitivity with respect to the wavenumber $\beta$ in the periodic direction becomes 
\begin{equation}
    \frac{1}{\sigma_j^2}\frac{\partial \sigma_j}{\partial \beta}
    \approx 
    \text{Real}\left\{\left<\hat{\mathbf{f}}_j,
    (i\mathcal{L}_1+2\beta \mathcal{L}_2)\hat{\mathbf{q}}_j\right>_E\right\}.
\end{equation}
Using the gradients, even with a coarse discretization of the parameter space, it is possible to map out the gains distribution. The gradients can be used in a cubic spline to interpolate the gain distribution between computed gain values. This can reduce computing time by avoiding a large number of SVDs, which would have been required if a finely discretized parametric space were to be adopted.

\section{Computational approaches}
\label{sec:Methods}

For a large resolvent operator $\mathcal{H}$, performing the singular value decomposition is numerically taxing.  This calls for efficient numerical algorithms to compute the leading resolvent modes in an accurate manner.  In this section, we discuss some approaches that can be taken to find the resolvent modes and gain for large-scale problems.

First, let us discuss some mathematical characteristics of the resolvent formalism.  Given the singular value decomposition introduced in Eq.~\ref{eq_2}, the following algebraic relations holds:
\begin{equation}\label{eq:resolventEigenProb}
    \mathcal{H}_\omega\mathcal{H}_\omega^*
        = Q_\omega \Sigma\Sigma^* Q_\omega^*
    \quad \text{and} \quad
    \mathcal{H}_\omega^*\mathcal{H}_\omega
        = F_\omega \Sigma^*\Sigma F_\omega^*.
\end{equation}
Noticing that $\Sigma \Sigma^*$ and $\Sigma^* \Sigma$ are diagonal matrices with the square of singular values as their entries, we notice that these two equations correspond to eigenvalue decompositions of $\mathcal{H}_\omega \mathcal{H}_\omega^*$ and $\mathcal{H}_\omega^* \mathcal{H}_\omega$.
This tells us that the $j$th columns of $Q_\omega$ and $F_\omega$ are the $j$th eigenvectors of $\mathcal{H}_\omega\mathcal{H}_\omega^*$ and $\mathcal{H}_\omega^*\mathcal{H}_\omega$, respectively, with eigenvalues $\sigma_j^2$.
Based on these observations, we will consider below some numerical algorithms to determine the resolvent triplets $(\hat{\mathbf{q}}_j,\sigma_j,\hat{\mathbf{f}}_j)$ that are founded on both eigenvalue and singular value decomposition properties.


Here, we present some techniques including Krylov projection \cite{arnoldi1951principle,loiseau2019time} and the randomized numerical linear algebra \cite{halko2011finding,ribeiro2020randomized} to efficiently compute the resolvent modes.  These methods aim to identify the singular triplets with the highest singular values through the definition of a low-order subspace that holds the important action of the matrix. This process is performed by a series of matrix-vector multiplications. However, as the resolvent operator is defined within an inverse operation, these processes translate to solving linear systems, which is the most demanding process of resolvent analysis computation.  For this reason, we also discuss the iterative time-stepper approach that has shown how to reduce the computational cost associated with solving such linear systems \cite{monokrousos2010global, martini2021efficient, gomez2013matrix, loiseau2019time}.  To aid with the implementation, we offer pseudocodes for key algorithms presented below.  These peudocodes can be found in Appendix \ref{AppendixB}.

\subsection{Krylov projection}
\label{sec:Krylov}

\begin{figure}
\centering
\includegraphics[width=\textwidth]{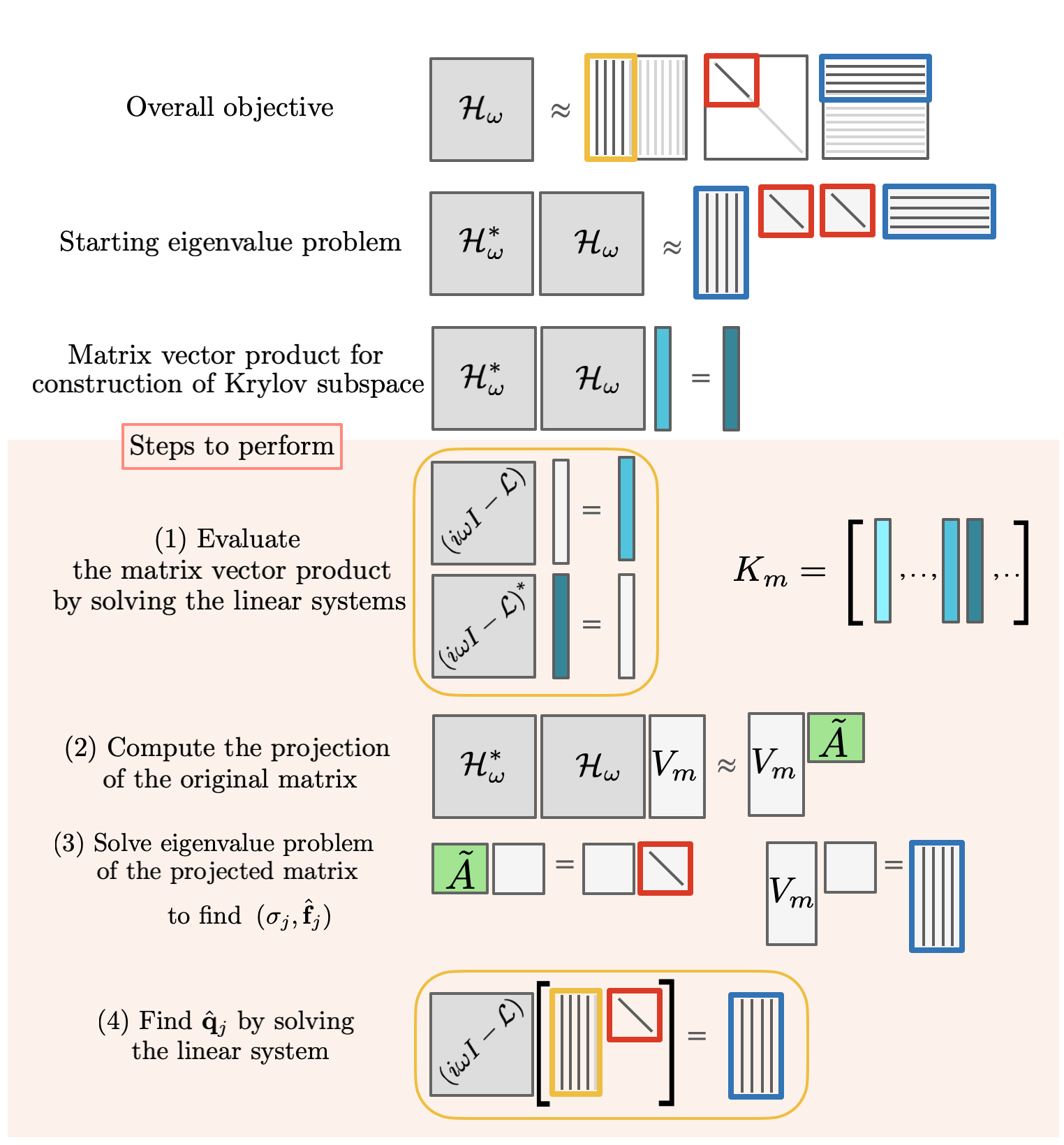}
\caption{\label{fig:SchemeKry}{Overview of the Krylov projection based resolvent analysis.}}
\end{figure}

A common approach to determine the resolvent triplet is to solve the eigenvalue problem (Eq. \ref{eq:resolventEigenProb} with respect to $\mathcal{H}_\omega\mathcal{H}_\omega^*$ or $\mathcal{H}_\omega^*\mathcal{H}_\omega$) \cite{arnoldi1951principle, stewart2002krylov, hernandez2007krylov, rolandi2021stability}. In this approach, the eigenvalues of a large-scale sparse matrix $A$ are approximated using Krylov projections by considering the problem in the Krylov subspace. 
We define the $m$-dimensional Krylov subspace of a matrix $A \in \mathbb{C}^{N \times N}$ as $K_m=\text{span}\{A^0\mathbf{x},A^1\mathbf{x},A^2\mathbf{x}, \dots ,A^{m-1}\mathbf{x}\}$, where $\mathbf{x}$ is generally a random initial vector.  The subspace converges toward the eigenvector associated with the eigenvalue with largest modulus and holds information on the leading eigenvalue and the corresponding eigenvector (i.e., eigenpair).  Considering $V_m\in  \mathbb{C}^{N \times m}$ an orthogonal basis of $K_m$, a low-order approximation of $A$ is formulated as
\begin{equation}\label{eq:projectionMat}
    AV_m\approx V_m \Tilde{A} ,
\end{equation}
where $\Tilde{A}$ is an upper Hessenberg matrix whose eigenpairs approximate those of $A$ \cite{loiseau2019time,hernandez2007krylov,rolandi2021stability}. Both the orthogonal basis $V_m$ and the projection matrix $\tilde{A}$ are computed through a Gram-Schmidt procedure, described in algorithm \ref{alg:ACompact}.

For resolvent analysis, the eigendecomposition of $A=\mathcal{H}_\omega^*\mathcal{H}_\omega$ is performed to obtain the singular triplet.  Recall that resolvent operator $\mathcal{H}_\omega=[-i\omega I-\mathcal{L}]^{-1}$ is defined with an inverse and is not generally available.  As mentioned previously, explicitly determining the inverse matrix is computationally impractical and inaccurate.  Therefore, we instead consider numerically solving the linear matrix equation
\begin{equation}
    [-i\omega I-\mathcal{L}][-i\omega I-\mathcal{L}]^*\mathbf{y} = 
    [-i\omega I-\mathcal{L}][i\omega I-\mathcal{L}^*]\mathbf{y} =\mathbf{x}
\end{equation}
to determine $A\mathbf{x}=\mathbf{y}$, which is needed for the generation of the Krylov subspace of a resolvent operator.
This is listed as step (1) in Fig. \ref{fig:SchemeKry} which shows the overall approach.
With this approach, we retrieve the eigenpairs $(\sigma_j^2,\hat{\mathbf{f}}_j)$ through steps (2) and (3) using the compact matrix $\tilde{A}$ as expressed in Eq.~\ref{eq:projectionMat} instead of having to deal with the full size $A$.

The left singular vector $\hat{\mathbf{q}}_j$ is instead computed considering the action of $\mathcal{H}_\omega$ on $\hat{\mathbf{f}}_j$, $\mathcal{H}_\omega \hat{\mathbf{f}}_j=\sigma_j \hat{\mathbf{q}}_j$, which is equivalent to solving the linear system
\begin{equation}
    [-i\omega I-\mathcal{L}]\hat{\mathbf{q}}_j=\frac{1}{\sigma_j} \hat{\mathbf{f}}_j,
\end{equation}
which is listed as step (4) in Fig. \ref{fig:SchemeKry}.  This procedure provides us with the complete triplet $(\hat{\mathbf{q}}_j,\sigma_j,\hat{\mathbf{f}}_j)$.

The cost of performing a Krylov projection method is $O(N^2m)$ with $m$ the dimension of the Krylov subspace, $N$ the dimension of the matrix and $m\ll N$. Also, the convergence of the modes is exponential with respect to the increase of the Krylov subspace dimension.

\subsection{Randomized resolvent analysis}
\label{sec:Randomized}

\begin{figure}
\centering
\includegraphics[width=\textwidth]{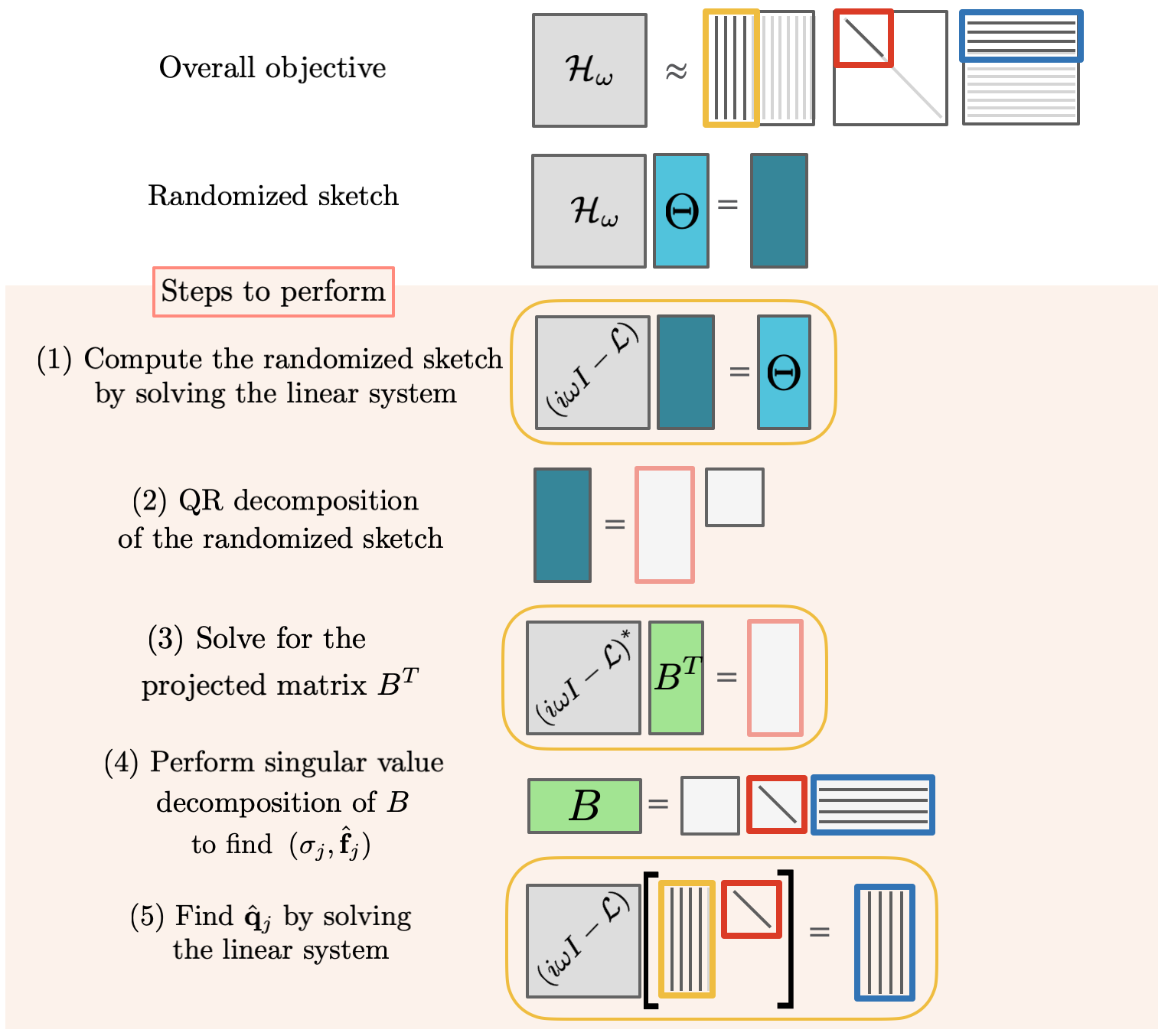}
\caption{\label{fig:SchemeRand}
Overview of the randomized resolvent analysis.} 
\end{figure}

The randomized resolvent analysis is a method based on a low-order approximation of the resolvent operator achieved through a projection of the original operator into an appropriate subspace.  First, let us consider a tall and skinny test matrix $\Theta \in \mathbb{R}^{N \times m}$ that is randomly generated.  The crux of this approach is that only a small number of test vectors $m \ll N$ are needed in $\Theta = [\boldsymbol{\theta}_1, ~\boldsymbol{\theta}_2, \dots, \boldsymbol{\theta}_m]$, which can make the SVD calculation lower in cost.  By passing this test matrix (or the collection of test vectors) through an operator $A \in \mathbb{C}^{N \times N}$, we can retain the dominant action of $A$ with 
\begin{equation}\label{eq:sketch}
    A \Theta = Y.
\end{equation}
The output $Y \in \mathbb{C}^{N \times m}$ is referred to as the sketch of $A$.
If we denote the $i$th column of the test matrix as $\boldsymbol{\theta}_i$, we observe that the sketch collects a sequence
\begin{equation}
    \{A \boldsymbol{\theta}_1,
    ~A \boldsymbol{\theta}_2,
    ~ \dots ,
    ~A \boldsymbol{\theta}_{m}\},
\end{equation}
which in a way resembles the Krylov sequence in the previous method.

If we consider $A=\mathcal{H}_\omega$, the computation of $Y$ corresponds to solving a linear system, which is step (1) in Fig.~\ref{fig:SchemeRand}. 
The orthogonalization of this $Y$ through a QR decomposition (step 2) creates a basis $\mathcal{Q}$ to derive the low-rank approximation of $\mathcal{H}_\omega$ as
\begin{equation}
    A\approx \mathcal{Q}\mathcal{Q}^*A=\mathcal{Q} B,
\end{equation}
where $B\in\mathbb{C}^{m\times N}$.  This again is determined by solving a linear system (step 3).  The process described until Step (3) is often named $\mathcal{Q}B$ decomposition, because it identifies the sketch basis $\mathcal{Q}$ and the low-order projection matrix $B$, in which the randomized SVD algorithms are founded. The algorithm to obtain the $\mathcal{Q}B$ decomposition of the resolvent operator, considering $A=[-i\omega I-\mathcal{L}]^{-1}$, is summarized in algorithm \ref{alg:qb}.

Note that energy norms are not considered in the pseudocodes presented herein. For cases where an energy norm is present, their influence is accounted for through matrix-vector multiplications performed before and after direct and adjoint linear systems are solved. After obtaining $\mathcal{Q}$ and $B$, the traditional and computationally less expensive manner to obtain the response modes $Q_\omega$ and the forcing modes $F_\omega$ requires the single SVD of $B$ and a matrix multiplication. 
Noteworthy here is that we obtain the complex-conjugate transpose of $B$, hence, $B \in \mathbb{C}^{m \times N}$ is a skinny row-matrix, upon which an SVD can be performed quickly for $m \ll N$. In this manner, through the SVD of the low-order projection matrix $B$, we can express
\begin{equation}
    A\approx \mathcal{Q}\Tilde{Q}_\omega\Sigma F_\omega^*.
\end{equation}
This approach follows the randomized SVD proposed in \cite{halko2011finding}, in which the singular triplet is identified as ($\mathcal{Q}Q_\omega$,$\Sigma$,$F_\omega$) shown in step (4) and described in algorithm \ref{alg:rsvd1}.

Taking advantage of the input-output characteristics of $A=[-i\omega I-\mathcal{L}]^{-1}$, a modification of the original algorithm was proposed \cite{ribeiro2020randomized}. Indeed, after computing the right singular vectors $\hat{\mathbf{f}}_j$ in step (4), it is possible to perform a further action of matrix $A$ in order to compute $\hat{\mathbf{q}}_j\sigma_j$ from $\hat{\mathbf{f}}_j$, which corresponds to step (5) in figure \cite{ribeiro2020randomized}. 
\begin{equation}\label{eq:sketch2}
    A\hat{\mathbf{f}}_j=\sigma_j \hat{\mathbf{q}}_j.
\end{equation}
Knowing that $\hat{\mathbf{q}}_j$ is a unit norm vector, we can scale the above results to find
\begin{equation}
    \hat{\mathbf{q}}_j=\frac{A\hat{\mathbf{f}}_j}{||A\hat{\mathbf{f}}_j||},\;\;\;\;\; \sigma_j=||A\hat{\mathbf{f}}_j||.
\end{equation}
This process has an added cost of solving an additional linear system, as summarized in algorithm \ref{alg:rsvd2}. In spite of the added computational cost, this approach results in more accurate computation of response modes and gains. This additional cost can be minor when the LU factorization of $A$ has been attained in the previous steps.  For iterative solvers, using the sketch $Y$ to initiate the solver is helpful to reduce computational time.

The aforementioned methods can improve the accuracy of resolvent gains and response modes, but have no effect on the forcing modes. In fact, for the previous approaches, the accuracy of forcing modes relies solely on the accurate computation of the low-order projection $B$ and its SVD. To obtain further accurate forcing modes, \cite{ribeiro2020randomized} provided an optional step with an additional SVD to improve the accuracy of forcing modes, as shown in algorithm \ref{alg:rsvd3}.

The cost of performing randomized resolvent analysis is $O(N^2 k)$ with $k \ll N$ being the desired number of singular values to compute. The convergence of the singular values with respect to an increasing number of the test vectors ($k$) is linear. Additionally, it is important to note that with the randomized method, each action on the test vectors can be computed independently, which can help decrease the computational cost of the operation.

\subsection{Time-steppers for projection}

The computational cost of handling large-sized operators in resolvent analysis can be considerably high, which has been one of the major challenges for applying resolvent analysis to high-Reynolds number turbulent flows.  A large computational grid is necessary for such flows, making direct computation of resolvent modes challenging. The major bottleneck is associated with the time and memory requirements for the linear systems solvers as part of the singular value decomposition in resolvent analysis.

\begin{figure}
\centering
\begin{tikzpicture}
\node[anchor=south west,inner sep=0] (image) at (0,0) {\includegraphics[page=1,trim=0mm 0mm 0mm 0mm, clip,width=0.5\textwidth]{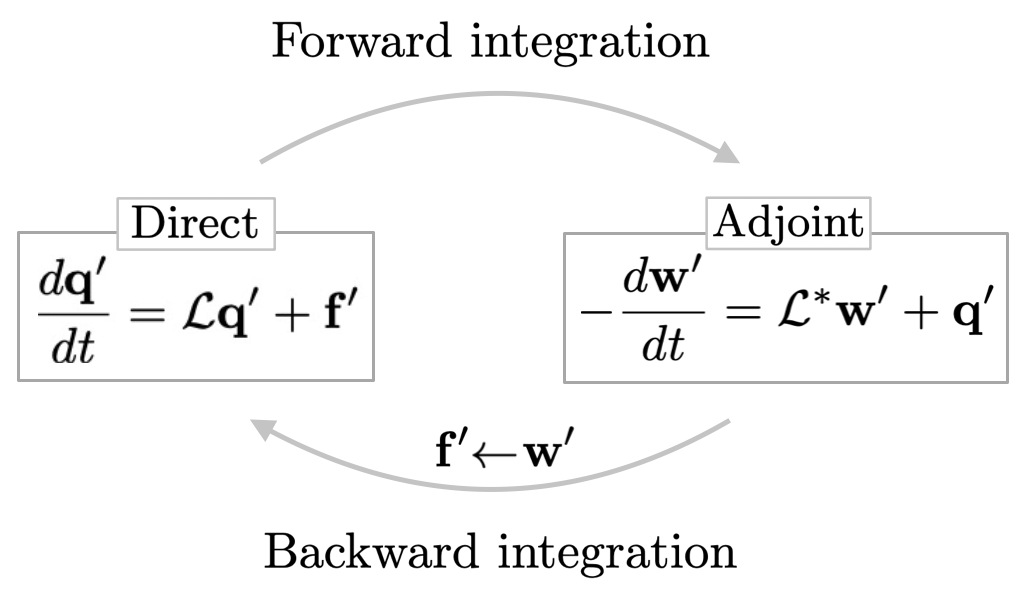}};
 \end{tikzpicture}
\caption{\label{fig:DirectAdjoint}Direct-adjoint integration loop for the solution of the optimization problem.}
\end{figure}

A n\"aive approach is to use direct linear system solvers based on LU decomposition. This approach, however, can be computationally taxing as the resolvent operator size increases. There are strategies to avoid the solution of the linear systems in homogeneous flows \citep{Barthel:PRF22}. Furthermore, iterative solvers and time-steppers have shown a potential to replace the direct solvers with a substantial reduction of memory requirements \citep{Barkley:IJNMF08,Monokrousos:JFM10,Gomez:JFM16}. The computational time for convergence may be high using traditional methods, but novel approaches have shown how to reduce it by using streaming discrete Fourier transforms \citep{Martini:JFM21,Farghadan:AIAA21}.

Time-steppers may be developed in matrix-free forms that need no explicit representation of the operator $\mathcal{L}$,  considerably reducing the storage needed for high-dimensional linear systems.  It is possible to approximate the effect of the resolvent operator by time integrating both the linearized and adjoint equations. Indeed, the step of solving the linear system, which is performed for each frequency $\omega$ for computing the iterative action of $\mathcal{H}_\omega^* \mathcal{H}_\omega$ on a vector for the construction of the Krylov sequence or the sketch basis in the randomized resolvent algorithms, is replaced by iterative direct-adjoint temporal integrations \cite{monokrousos2010global, martini2021efficient}.

The direct and the corresponding adjoint equations read as follow
\begin{equation}\label{eq:SystDirAdj}
    \frac{d\mathbf{q}'}{dt} = \mathcal{L} \mathbf{q}' + \mathbf{f}'
    \quad \text{and} \quad
    -\frac{d\mathbf{w}'}{dt} = \mathcal{L}^* \mathbf{w}' + \mathbf{q}'.
\end{equation}
This direct-adjoint action in frequency space translates to 
\begin{equation}\label{eq:AdjDir}
    \hat{\mathbf{w}}=\mathcal{H}_\omega^*\mathcal{H}_\omega \hat{\mathbf{f}}.
\end{equation}
This means that the direct-adjoint integration, depicted in Fig. \ref{fig:DirectAdjoint}, serves as an approximation for the action of $\mathcal{H}_\omega^*\mathcal{H}_\omega$. 
This can be used for the construction of the Krylov sequence and also for the direct and adjoint linear systems solvers for the randomized algorithm.

For example, the optimal modes at the forcing frequency can be retrieved by starting the integration loop with a harmonic real-valued or complex-valued forcing, and using the power iteration method ($\lim_{m\to\infty}K_m$) \cite{monokrousos2010global,gomez2016estimation}. Then, \cite{martini2021efficient} proposed two variations of this method for computing $\hat{\mathbf{w}}$ in Eq.~\ref{eq:AdjDir}: the transient-response method (TRM), and the steady state response solution (SSRM). TRM computes the solutions of the integration loop providing a compact forcing on time. Therefore, TMR computes the transitional input-output response, while the SSRM computes the steady asymptotic input-output response of the continuously forced system. Moreover, with these two approaches they also proposed to use the full response and forcing variables for computing the input-output response at each frequency simultaneously. Recently, \cite{farghadan2023scalable} proposed another method, called the RSVD-$\Delta t$, which couples the randomized resolvent analysis and the steady state response approach \cite{martini2021efficient}, showing reduced computational cost and memory consumption which allows to perform resolvent analysis of three-dimensional flows in a very efficient way.

Indeed, with the temporal integration loop, we obtain 
\begin{equation}
    \mathbf{w}^\prime(\mathbf{x},t)=\int_{-\infty} ^{\infty}\hat{\mathbf{w}}(\mathbf{x})e^{-i\omega t}d\omega.
\end{equation}
Therefore, the Krylov and sketch sequences can be composed for each $\omega$ from the Fourier transforms of the signal $\mathbf{w}^\prime(\mathbf{x},t)$ allowing for the evaluation of the input-output response with respect to $\omega$.\\

For these methods presented, the choice of the initial vector for the Krylov projection methods, or the test vectors for the randomized resolvent plays a very important role. Indeed, initial vectors that are closer to the relevant subspace yield a considerable reduction in the demanded size of the low-order subspace for accurate computation of resolvent modes. 
Realizing that the response and forcing modes mainly appear in regions of high gradients (shear) in the base flow, weighting the initial random vector (for the Krylov projection methods) and the test vectors (for the randomized resolvent) with the base flow gradients achieves faster convergence \citep{ribeiro2020randomized}.

\subsection{Data-driven methods}

For linearly stable flows, resolvent modes may be derived from data without access to the adjoint simulations \cite{Herrmann:JFM2021datadriven}. Through this data-driven resolvent analysis, forcing and response modes can be found in an equation-free manner, without the need for adjoint simulations. The foundation of this method is the dynamic mode decomposition \cite{SchmidJFM2010dmd,Schmid:ARFM2022dmd,Kutz2016book}, which identifies the linearized flow dynamics based on snapshots obtained from a collection of transient trajectories of the flow. 

First, consider a set of $m$ snapshots for $p$ distinct flow trajectories stored within the data matrices $X$ and $Y$ as
\begin{eqnarray}
    X &=&[\mathbf{x}_1^{(1)} ~\dots~ \mathbf{x}_{m-1}^{(1)} 
       ~\dots~ \mathbf{x}_1^{(p)} ~\dots~ \mathbf{x}_{m-1}^{(p)}],\\
    Y &=&[\mathbf{x}_2^{(1)} ~\dots~ \mathbf{x}_m^{(1)} 
       ~\dots~ \mathbf{x}_2^{(p)} ~\dots~ \mathbf{x}_m^{(p)}].
\end{eqnarray}
Here, the snapshots in $X$ and $Y$ are shifted by a time index, which amounts to a time step of $\Delta t$.
Given these data matrices, the time step, and the desired number of terms $r$, the dynamic mode decomposition is performed to obtain
\begin{equation}
    \text{DMD}(X,Y,\Delta t, r) \rightarrow \Lambda_r, V_r, U_r,
\end{equation}
where the reduced-order dynamics of the flow system is identified through the eigenvalues presented in the diagonal of $\Lambda_r$, along with the direct and adjoint eigenvectors, $V_r$ and $U_r$, respectively. 

The $r$ eigendecomposition triplets are then used to build a projected resolvent operator
\begin{equation}
    \tilde{\mathcal{H}}_\omega = (-i\omega I- \Lambda_r)^{-1}.
\end{equation}
Here, $\tilde{\mathcal{H}}_\omega$ holds the linear dynamics through the eigenvalues of the dynamic mode decomposition. The projected resolvent modes are then identified through SVD as
\begin{equation}
    \tilde{\mathcal{H}}_\omega 
    = \tilde{\Psi}_\omega \tilde{\Sigma}_\omega \tilde{\Phi}_\omega
\end{equation}
in which $\tilde{\Sigma}_\omega$ is the data-driven resolvent gain, while  $\tilde{\Psi}_\omega$ and $\tilde{\Phi}_\omega$ are the projected forcing and response modes. The final step needed for data-driven resolvent modes is to synthesize them in physical coordinates with
\begin{eqnarray}
    \Phi = V_r \tilde{\Phi} 
    \quad \text{and} \quad 
    \Psi = V_r \tilde{\Psi}.
\end{eqnarray}
A weighting norm can be used in the data-driven resolvent analysis to account for appropriate scaling \cite{Herrmann:IJHMT2018heat}. In such case, a positive definite weighting matrix $F$ must be defined through the Cholesky decomposition of a physically meaningful positive definite $W = F^* F$ analogous to those described in Sec.~\ref{sec:energyNorm}.

The data-driven resolvent analysis has been shown to find resolvent modes for low Reynolds number flows \cite{Herrmann:JFM2021datadriven}.  Future efforts are hoped to enable its applications to higher Reynolds number problems, and possibly to experimental data sets.  For a detailed description of the data-driven resolvent analysis, we point to reference \cite{Herrmann:JFM2021datadriven}.

\section{Case studies}
\label{sec:StepByStep}

In this section, we outline the step-by-step procedure for conducting resolvent analysis and provide two external flow examples.  First, we discuss the preparations to perform resolvent analysis, commenting on the computational domain size and providing insights into strategies for achieving reduction in computational cost and memory usage.  Second, we present examples of both biglobal and triglobal resolvent analysis. The first case study is a biglobal resolvent analysis of a separated flow around a NACA0012 airfoil. Following this, we discuss a triglobal resolvent analysis applied to the flow around a finite wing. Finally, we explore the use of resolvent analysis in flow control applications.

\begin{figure}
\centering
\includegraphics[width=\textwidth]{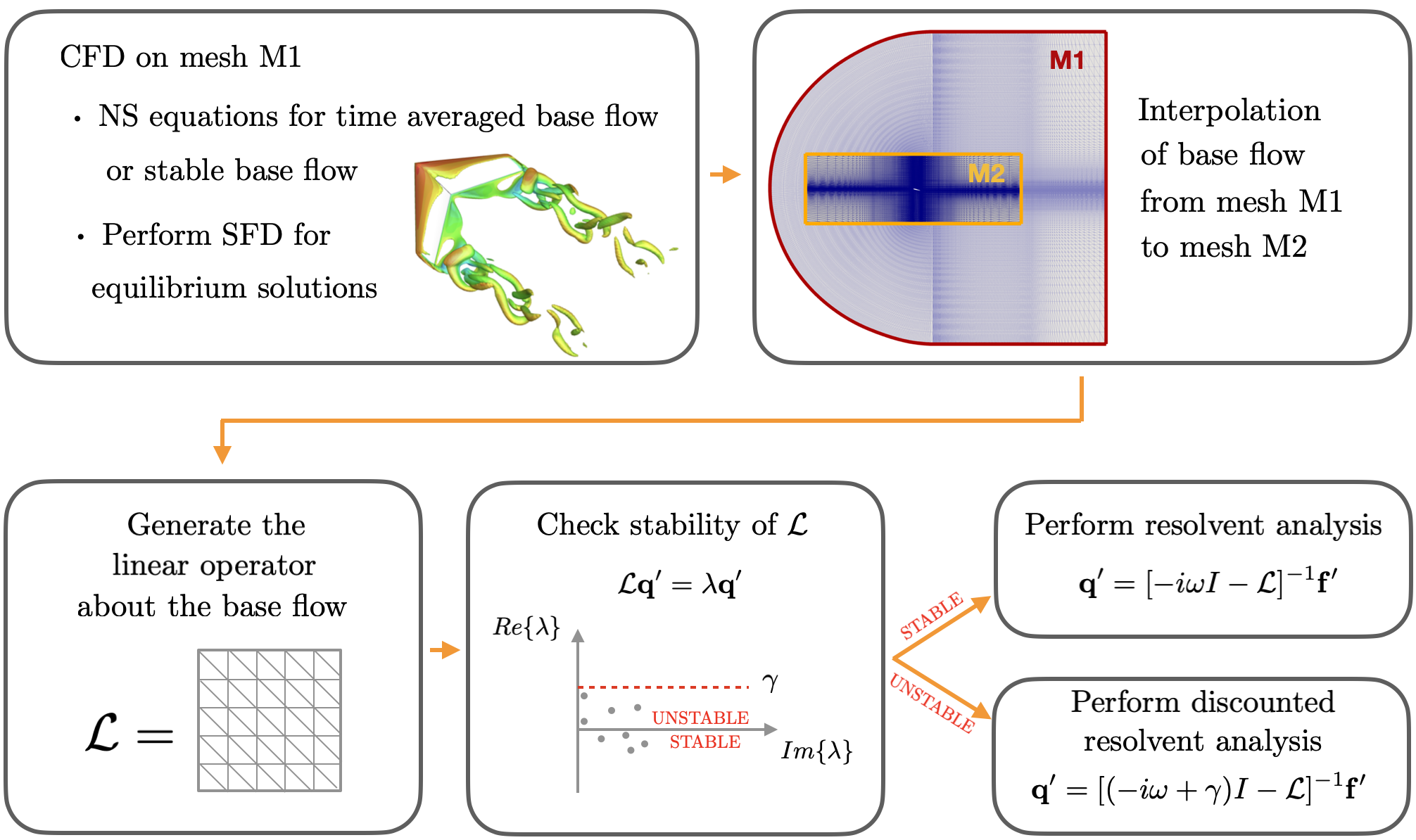}
\caption{\label{fig:DiagramResolvent}Steps for performing the resolvent analysis using a base flow from CFD.}
\end{figure}

\subsection{Preparation}\label{sec:set-up}

As we prepare to perform resolvent analysis, let us go over the necessary steps, as shown in Fig. \ref{fig:DiagramResolvent}. 
While we try to keep the discussions as general as possible, some of them may be specific to external flow problems.   
The first step of performing the stability analysis and resolvent analysis concerns the choice of the base flow.  We need to consider whether an equilibrium state can be found or if the time-averaged base flow is to be used.  In these cases, it is important that the stability of the base flow is analyzed to see if we need to use discounting or an eddy-viscosity model to provide a valid perspective.  We emphasize that the base flow should be accurate and reliable as the rest of the resolvent analysis depends on the fidelity of its discretizations.  The computational domain needed for the base flow calculation should follow the classical guidelines of LES/DNS \cite{KajishimaTaira} using spatial and temporal discretization schemes with far-field boundary conditions placed sufficiently away from the body to obtain an accurate base flow.  This base flow should of course be verified and validated carefully.

The next step is to construct a linear operator $\mathcal{L}$ about the chosen base flow.  For many practical reasons, an optional step of interpolating the base flow onto a different mesh can be computationally beneficial, or even necessary, on which the linear operator is to be discretized. 

\begin{figure}
\centering
\includegraphics[width=\textwidth]{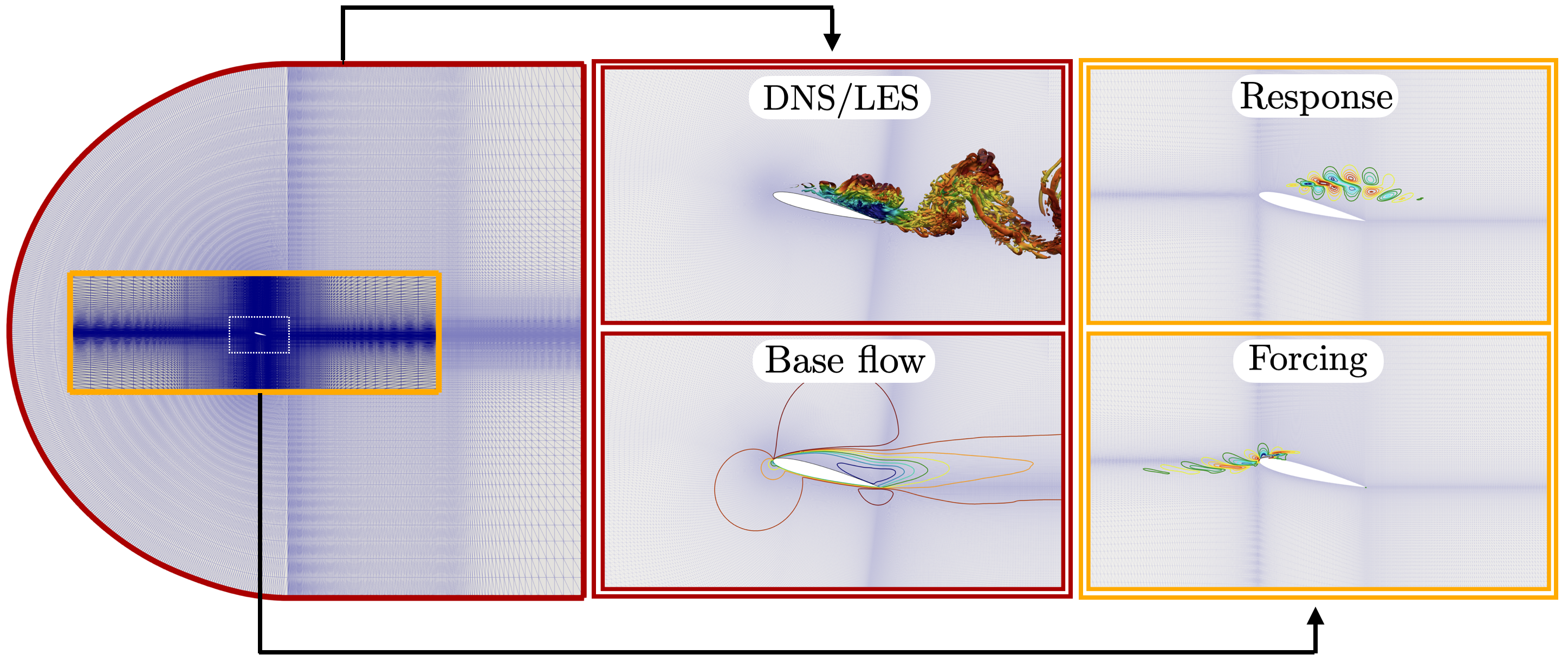}
\caption{\label{fig:MeshDetails} Meshes for computing the base flow (red) and the resolvent analysis (yellow) are shown on the left. Zoom-in views are shown in the center and right around the airfoil for base flow and resolvent mode computations. The shown example is for flow around a NACA0012 airfoil at angle of attack $\alpha=14^\circ$ and Reynolds number $Re=10,\!000$.}
\end{figure}

The first and rather obvious reason for choosing a different mesh is to reduce the computational cost associated with the solution of the linear system (action of the matrix inverse) and the singular value decomposition.  Since frequency is a parameter of choice in resolvent analysis, the associated length scale can usually be estimated in accordance with relevant advective (or diffusive) physics.  The knowledge of such a length scale can be leveraged to prepare a resolvent mesh whose spatial resolution is carefully tailored according to predetermined length scales.  This can lead to a significantly reduced grid size from that of the CFD mesh.  While a large computational domains is required for CFD to allow for the free-stream condition to be fully recovered, a much smaller may be used for resolvent analysis as the fluctuations decay to almost zero much closer to the body.  This allows for a smaller computational domain to be used in resolvent analysis of external flows. Therefore, instead of using the original CFD mesh that resolves turbulent length scales as small as possible, a coarser mesh with a smaller domain extent can usually be adapted in the resolvent analysis to reduce the cost of involved numerical linear algebra.  This is demonstrated in Fig. \ref{fig:MeshDetails}, where the CFD mesh used for the computation of base flow is highlighted in red, and the coarser resolvent mesh with a smaller domain is highlighted in yellow.  We caution the readers that if a different mesh is to be chosen, the results from resolvent analysis should be checked for mesh independence, just as the base flow verification is performed.

The second and important reason for using a resolvent-specific mesh concerns the forcing/adjoint mode, whose spatial support (where they appear in space) can be significantly different from that of the turbulent structures captured in direct CFD simulations.
As shown in Fig.~\ref{fig:MeshDetails}, the forcing mode for external flows emerges upstream of the airfoil. Therefore, the mesh necessitates sufficient spatial resolution in that region in order to resolve the forcing mode structure.  The region upstream of a body does not usually have a very refined grid for direct CFD simulation and a relatively aggressive grid stretching may be applied.  This is indeed the case for the example in Fig.~\ref{fig:MeshDetails}, where CFD is performed on a C-grid where aggressive grid stretching in the upstream of the airfoil would result in inaccurate recovery of forcing mode and resolvent gain.
In such case, interpolating the base flow onto a resolvent-specific mesh becomes necessary in order to obtain the forcing modes with sufficient spatial resolution.  This is demonstrated in Fig.~\ref{fig:MeshDetails}, where the base flow obtained from the C-grid is interpolated onto an H-grid that has the same levels of spatial resolution both upstream and downstream of the body.

Once the linear operator $\mathcal{L}$ about the chosen base flow is prepared, we solve the eigenvalue problem, Eq.~\ref{eq:GSA}, to obtain the spectrum of $\mathcal{L}$ near the stability margin. In this analysis, the main focus is not the linear stability of the flow.  In fact, for base flows that do not represent an equilibrium point, linear stability results must be taken with care. Our goal is to identify the eigenvalues of $\mathcal{L}$ with higher positive growth rate to set an appropriate discounting parameter for resolvent analysis (see Sect. \ref{sec:discounted}).

Below, we introduce two specific examples of performing biglobal and triglobal resolvent analysis and giving insights on the interpretation of the results. In these examples, both the calculation of the base flow and resolvent analysis are performed with the compressible flow solver CharLES \citep{khalighi2011unstructured}, coupled with the PETSc and SLEPc libraries \citep{balay2020petsc,roman2016slepc} for performing the singular value decomposition. We specifically use the Krylov-Schur projection method and the randomized resolvent for the singular value decomposition. An LU factorization of the linear operator, instead, is used for solving the linear system, needed for the construction of the Krylov subspace and the sketched matrix. Moreover, for the three-dimensional case, the MUMPS (multifrontal massively parallel sparse direct solver) package \citep{Amestoy:SIAM01} is also used within PETSc, in order to parallelize the LU factorization and singular value decomposition.  In the triglobal example, the codes used to compute the resolvent modes are part of the `linear analysis package' made available by \cite{SkeneRibeiro:tools}.
Towards the end of this section, we also briefly discuss how the insights from resolvent analysis can be used to guide flow control designs.

\subsection{Biglobal analysis}

\begin{figure}
\centering
\includegraphics[width=\textwidth]{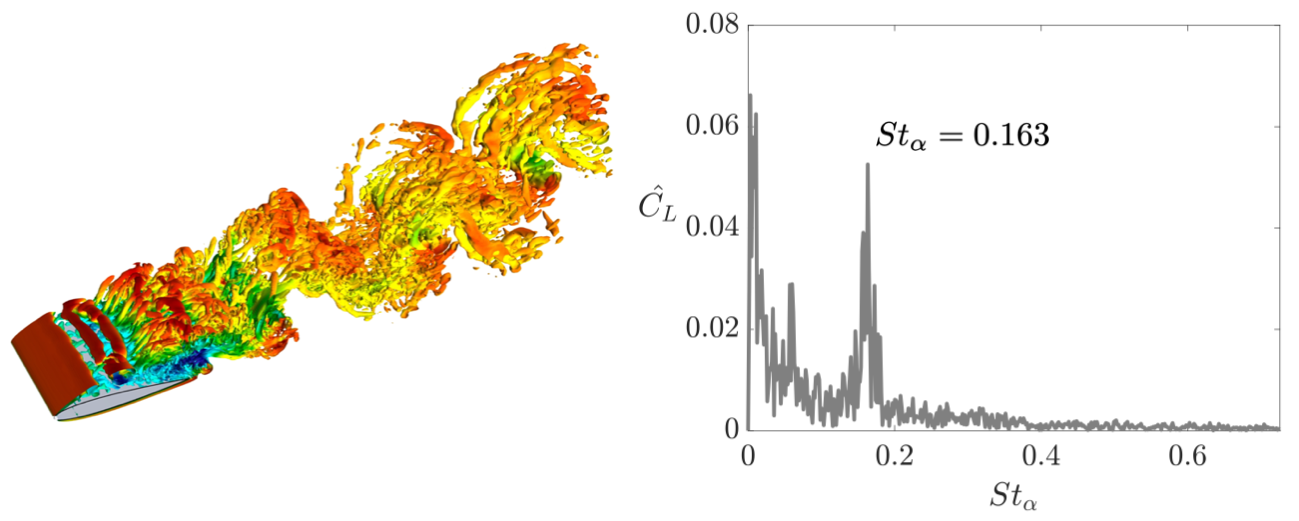}
\put(-370,150){(a)}
\put(-190,150){(b)}
\caption{\label{fig:BaseFlowBig} Spanwise periodic turbulent flow over a NACA0012 airfoil at $\alpha=14^\circ$ and $Re=10,\!000$. (a) Isocontour of Q-criterion of the instantaneous flow field colored by the streamwise velocity and (b) lift coefficient spectra ($\hat{C}_L$) over frequency ($St_{\alpha}$).}
\end{figure}

Let us consider a separated turbulent flow around a two-dimensional airfoil at an angle of attack $\alpha=14^\circ$ and chord based Reynolds number $Re=10,\!000$. The instantaneous flow field is shown in Fig.~\ref{fig:BaseFlowBig}(a) and the lift coefficient spectra reveals a characteristic von Karman shedding frequency at $St_\alpha={fc\sin{\alpha}}/{U_\infty}=0.163$, as reported in Fig.~ \ref{fig:BaseFlowBig}(b). Here, $c$ is the chord, $f$ is the frequency and $U_\infty$ is freestream velocity. At this Reynolds number, the flow is three-dimensional.  To computing the base flow, LES is performed over a spanwise periodic domain whose spanwise extent captures the three-dimensional flow structures. A spanwise length of $L_z=c$ has been chosen for this calculation. The base flow is obtained by averaging the flow over time and the spanwise direction. Therefore, the base flow is a two-dimensional base flow (homogeneous along the spanwise direction) and a biglobal formulation is used for constructing the linear operator $\mathcal{L}=\mathcal{L}(\beta)$, which depends on the spanwise wavenumber $\beta$.  Details on the formulation can be found in Appendix \ref{AppendixA}).

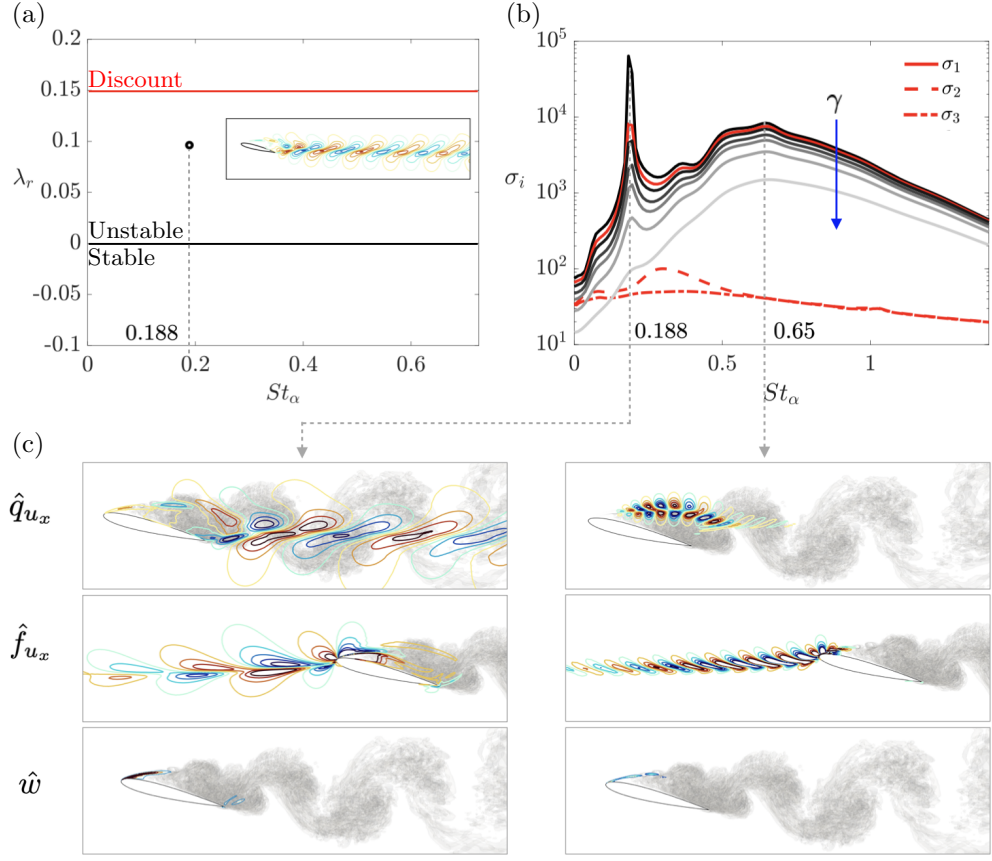
\begin{figure}
\centering
\begin{tikzpicture}
\node[anchor=south west,inner sep=0] (image) at (0,0) {\includegraphics[page=1,trim=0mm 0mm 0mm 0mm, clip,width=1\textwidth]{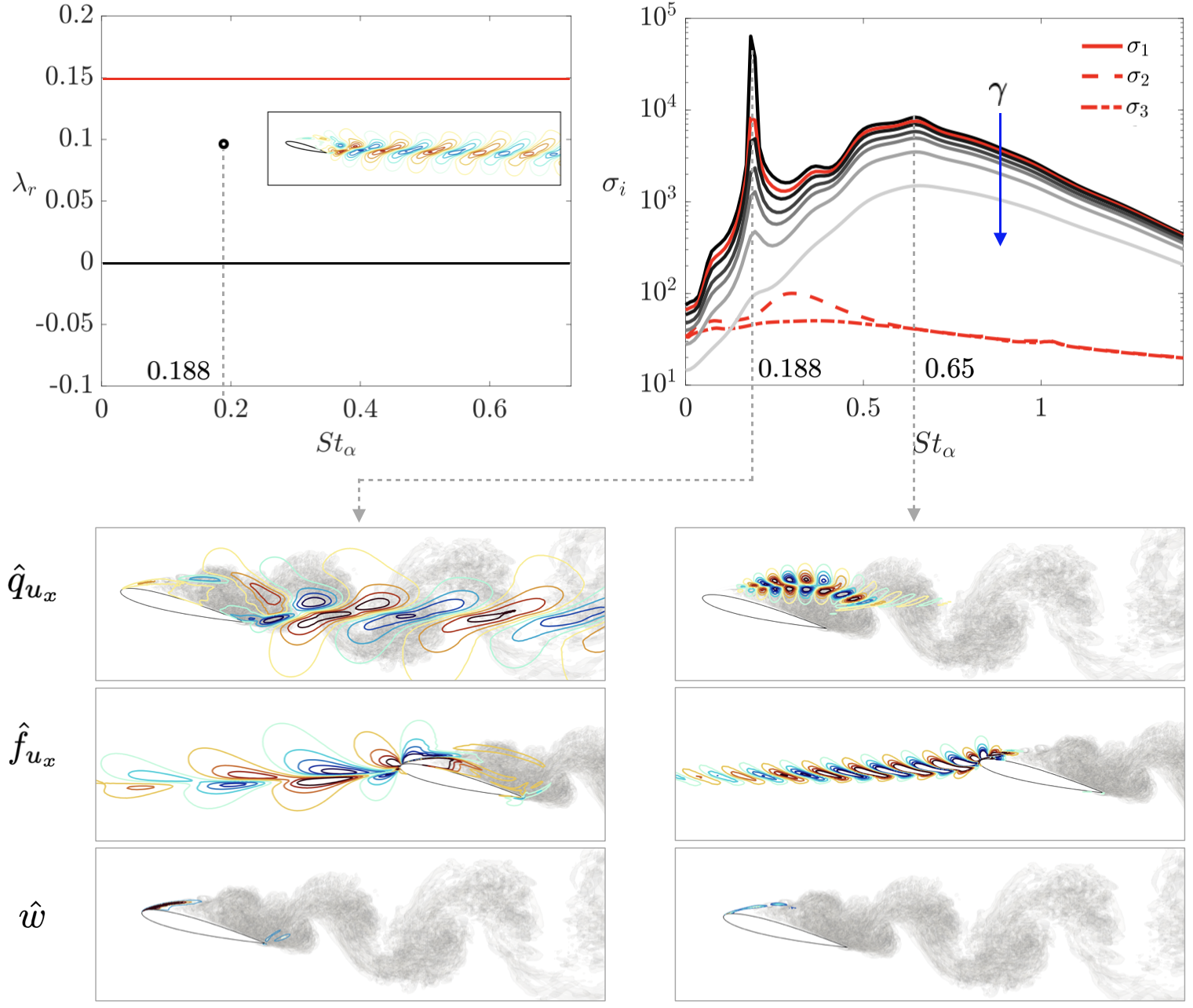}};
\node[anchor=west] at (1,10.35) {\small{{\color{red}Discount}}};
\node[anchor=west] at (1,8.35) {\small{Unstable}};
\node[anchor=west] at (1,8) {\small{Stable}};
\node[anchor=west] at (0.,11.2) {(a)};
\node[anchor=west] at (6.5,11.2) {(b)};
\node[anchor=west] at (0.,5.5) {(c)};
\end{tikzpicture}
\caption{\label{fig:DiscountGain}Biglobal resolvent analysis of the NACA0012 airfoil wake. (a) The most unstable eigenvalue and the corresponding eigenvector visualized with isocontours of streamwise velocity. (b) The first three dominant gains (singular values $\sigma_1$, $\sigma_2$ and $\sigma_3$) over $St_\alpha$ shown in red. Also presented is the influence of discount parameter $\gamma$ on $\sigma_1$ illustrated in grayscale. The red curves use the $\gamma$ value from (a). (c) Response and forcing modes along with the wavemaker for $\gamma=0.15$ at $St_\alpha=0.188$ and $0.65$. Visualization of isocontours of streamwise velocity components for $\hat{f}_x$ and $\hat{q}_x$ and isocontours of $\hat{w}$ on top of the instantaneous flow field (Q-criterion).}
\end{figure}

In this example, the least stable eigenvalue has a positive growth rate $\lambda_r=\text{Real}(\lambda)=0.096$, as shown in figure \ref{fig:DiscountGain}(a). The system is thus unstable calling for a discount parameter $\gamma >\lambda_r$ in the resolvent operator (see Sect. \ref{sec:discounted})
\begin{equation}
    \mathcal{H}_\omega= [(-i\omega + \gamma)I -\mathcal{L}(\beta) ]^{-1}.
\end{equation}
We perform singular value decomposition of $\mathcal{H}_\omega$ over $\omega  \in [\omega_\text{init},\omega_\text{end}]$ with $\Delta\omega=(\omega_\text{end}-\omega_\text{init})/N_{\omega}$. In this case, we used $\Delta \omega=(6.2\times 2\pi)/50$ giving a frequency resolution of $\Delta St_\alpha = \Delta  \omega \sin{\alpha}/2\pi =1.5/50$. During the singular value decomposition, we also evaluate the gradients of the gain with respect to $\omega$ to enable accurate interpolation, as discussed in Sect.~\ref{sec:Gradient}.  For the singular value decomposition, the Krylov-Schur projection method is used (Sect. \ref{sec:Krylov}) with a dimension of the Krylov subspace $m=30$.  In this examination, the first three singular triplets are sought.

Let us present the gains $\sigma_i$ over $St_\alpha$ in Fig. \ref{fig:DiscountGain}(b) for a spanwise wavenumber $\beta=0$ (2D mode) for the varied discount parameter of $0.1 \le \gamma \le 1.2$.  The primary gain $\sigma_1$ obtained with $\gamma=0.15$ is shown in red as a reference and reported together with second and third gains, $\sigma_2$ and $\sigma_3$, respectively.  The higher order gains are smaller than $\sigma_1$ by two orders of magnitude, suggesting that the primary resolvent triplet can form an accurate approximation of the resolvent operator
\begin{equation}
    \mathcal{H}_\omega
    \approx
    \hat{\mathbf{q}}_1\sigma_1\hat{\mathbf{f}}_1^*.
\end{equation}
This particular expression is referred to as the rank-1 approximation.

With this choice of discount parameter, we are effectively considering the input-output relation over a finite-time horizon of $t_\gamma = 2\pi/0.15 = 41.9$. We can observe that varying the horizon time, the gain variation exhibits two different peaks.  The peak at $St_\alpha= 0.65$ is associated with the oscillations in the separated shear layer over the airfoil, as captured in the response modes in Fig.~\ref{fig:DiscountGain}(c). This higher frequency mode captures the dominant fast oscillation in the flow.  
While not shown in this paper, it is possible to visualize how $\hat{\mathbf{q}}$ and $\hat{\mathbf{f}}$ oscillate over time.  Recalling Eq.~\ref{eq:FourierTrans}, the oscillations of these modes can be expressed as $\hat{\mathbf{q}}e^{-i\omega t}$ and $\hat{\mathbf{f}}e^{-i\omega t}$ over the oscillation period $T=2\pi/\omega$.  

On the slower time scale, a distinct peak at $St_\alpha= 0.188$ can be identified. This frequency content of the flow acts in contrast in the wake as highlighted by the response mode structure in Fig.~\ref{fig:DiscountGain}(c).  This frequency agrees with the wake instability revealed by the stability analysis (eigenvalues with frequency at $St_\alpha= 0.188$).  In fact, the relationship between $\sigma_1$ and $\lambda$ can be explained with pseudospectral theory \cite{trefethen1993hydrodynamic, trefethen_pseudospectra_book}.

The shift of the response mode from the wake toward the shear-layer region is due to the fact that higher frequency corresponds to smaller characteristic length in the flow, which is predominantly supported by the shear-layer instability \citep{yeh2019resolvent}.
The fact that the gain peak at short times is at $St_\alpha= 0.65$ can be explained by the fact that shear regions are known to host nonmodal growths \citep{schmid2007nonmodal}.  Therefore, over short time, we observe higher gain from these mechanisms. However, the real parts of the eigenvalues associated with the shear-layer modes are negative.  In contrast, the real part of the eigenvalues for wake-region mode are positive suggesting comparatively dominant behavior. In Fig.~\ref{fig:DiscountGain}(c) we also report the wavemaker, which highlights regions in the shear layer associated with self-sustained mechanisms, see \citep{skene2022sparsifying,qadri2017frequency} for details.

\subsection{Triglobal analysis}

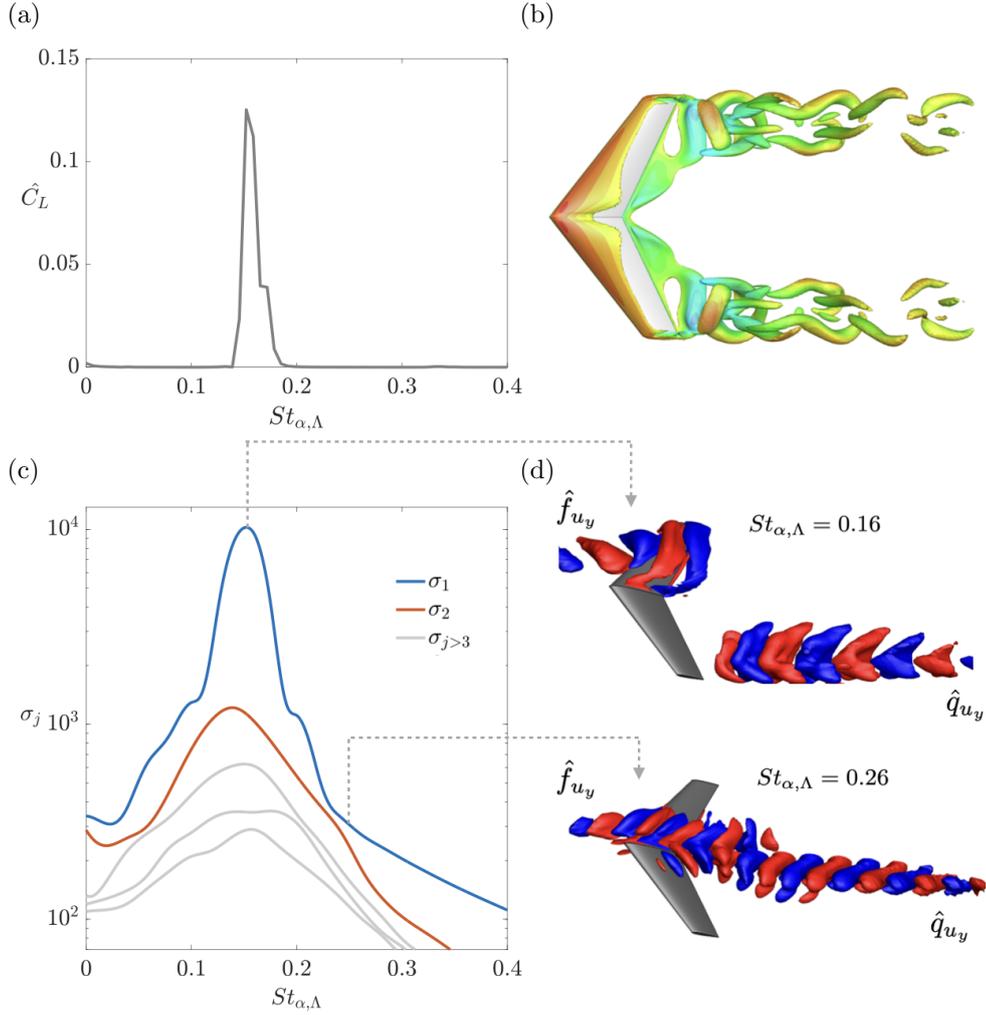
\begin{figure}
\centering
\includegraphics[width=\textwidth]{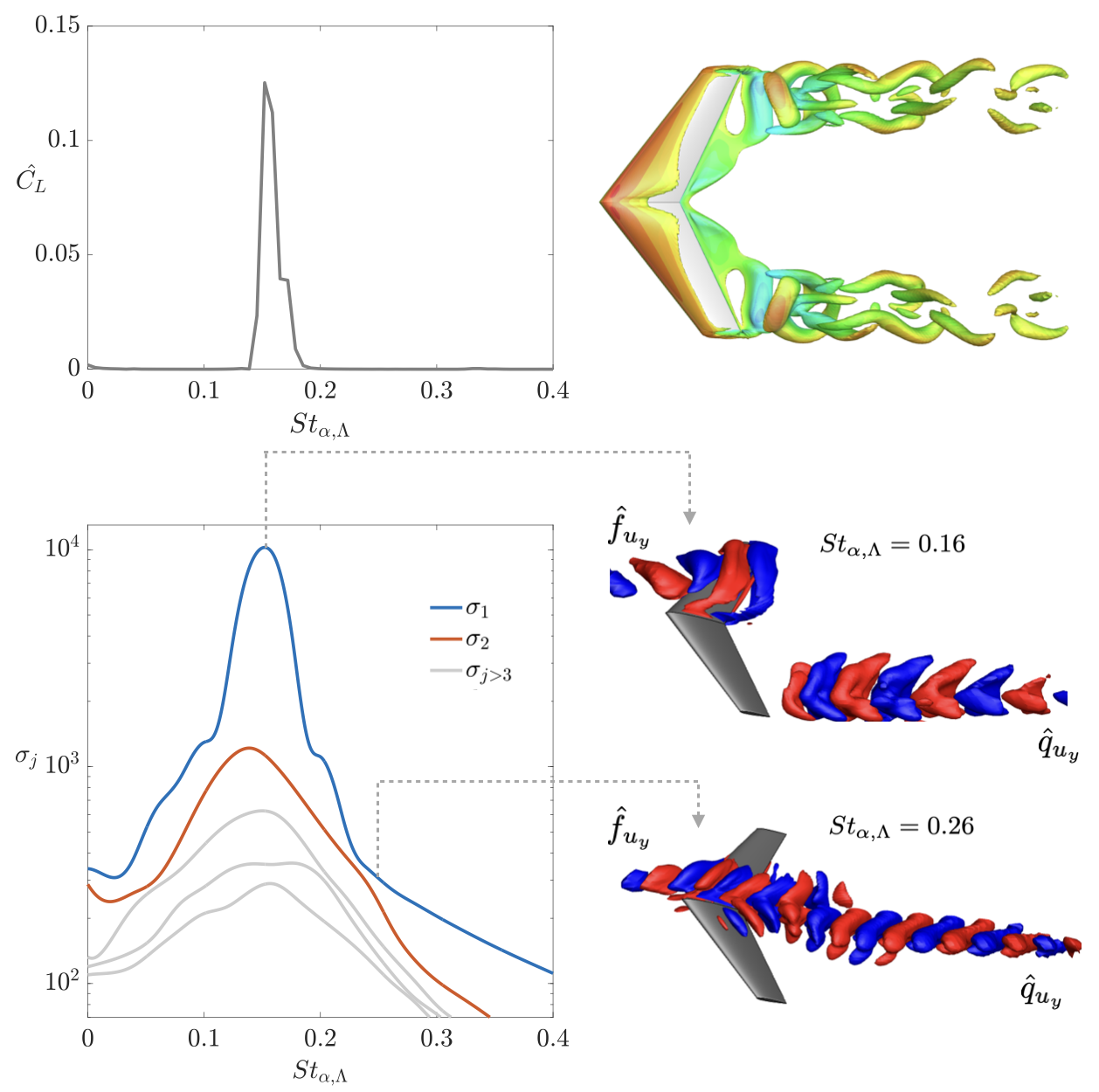}
\put(-372,375){(a)}
\put(-180,375){(b)}
\put(-372,203){(c)}
\put(-180,203){(d)}
\caption{Triglobal resolvent analysis of the flow over a tapered swept wing with semi aspect ratio $sAR = 2$, taper ratio $0.5$, leading edge sweep angle $\Lambda = 40^\circ$ at $Re=600$ and $\alpha = 22^\circ$. (a)  Lift coefficient ($\hat{C}_L$) spectra over frequency ($St_{\alpha,\Lambda}$). (b) Isocontour of Q-criterion of the instantaneous flow field colored by streamwise velocity. $(c)$ Resolvent gains ($\sigma_j$) over frequency ($St_{\alpha,\Lambda}$). $(d)$ The primary forcing (top half) and response (bottom half) modes are visualized with isosurfaces of cross-stream component at frequencies $St_{\alpha,\Lambda} = 0.16$ and $0.26$.}   \label{fig:triglobalResolventMode}
\end{figure}

A triglobal resolvent analysis is conducted to examine flows over finite-span wings.  The flow to be examined does not have any homogeneous directions, which necessitates a three-dimensional (triglobal) approach.  
The triglobal modes are characterized by three-dimensional global forcing and response modes \cite{Ribeiro:JFM23triglobal,Houtman:Flow23resolvent,Ribeiro:AIAA23resolvent}. 
In this example, we consider the flow over a tapered swept wing with semi-aspect ratio $sAR = 2$, taper ratio $0.5$, leading edge sweep angle $\Lambda = 40^\circ$ at $\alpha = 22^\circ$ and $Re = 600$ \cite{Ribeiro:JFM2023tapered}. This translating wing exhibits regions of steady and unsteady flows over the wingspan, as illustrated in Fig. \ref{fig:triglobalResolventMode}(b). Notably, the wing tip experiences significant flow unsteadiness and the formation of vortex shedding structures, while inboard flow remains primarily steady, characterized by a pair of large leading-edge vortices. The lift coefficient spectrum, shown in Fig.~\ref{fig:triglobalResolventMode}(a), indicates the characteristic shedding frequency at $St_{\alpha,\Lambda} = {fc\sin{\alpha}}/{(U_\infty \cos{\Lambda})} \approx 0.16$.  This Strouhal number takes the sweep angle into consideration.

Here, the singular value decomposition is performed with the randomized resolvent analysis (Sect. \ref{sec:Randomized}) with $m=10$ test vectors. For the resolvent analysis here, discounting is applied, whose effects share similarities with the aforementioned presentation for biglobal modes. In addition, similar to biglobal resolvent modes, triglobal modes are also non-uniform in space and exhibit larger amplitude over certain regions of the domain. These areas are characterized by a larger spatial support of forcing-response modes. In the streamwise $x$-direction, owing to the convective nature of the linear operator, forcing modes emerge upstream of the wing, while response modes are primarily exhibited downstream of the wing. As illustrated in Fig.  \ref{fig:triglobalResolventMode}(d), the forcing modes at $St_{\alpha,\Lambda} = 0.16$ emerge over and upstream of the wing, while the response modes appear downstream. Note that this mode pair is at the characteristic shedding frequency $St_{\alpha,\Lambda} = 0.16$ of this wing. The response modal structures align with the wing tip, exhibiting a larger spatial support over the same region where shedding structures are observed in the numerical simulations. 

For the triglobal analysis conducted over finite-span wings, the spatial support of forcing-response mode pairs is significantly influenced by the modal frequency. As illustrated in Fig.~\ref{fig:triglobalResolventMode}(d), as the frequency increases from $St_{\alpha,\Lambda} = 0.16$ to $0.26$, the larger spatial support transitions from the wing tip towards the wing root. Previous studies have observed a similar transition in triglobal analysis over untapered swept wings, where the modes transition from tip-based modes near the shedding frequency to root-based modes at higher frequencies. Conversely, for flows over untapered unswept wings, this trend is inverted. In such cases, forcing-response mode pairs transition from root-based to tip-based modes as the frequency increases \cite{ribeiro2023triglobal}. Generally, the dominant forcing-response mode pair at the frequency near the vortex shedding exhibits larger spatial support over the regions where flow unsteadiness occurs. As the frequency varies, modes switching shows unsteadiness in other regions of the flow. These findings are particularly relevant for flow control applications. By virtue of the input-output characteristics of the resolvent modes, the optimal location to perturb the flow for wake modification can be identified. 

The Triglobal resolvent analysis entails much higher computational costs compared to its biglobal counterpart. This is due to the generally larger size of the triglobal linear operator. However, even when considering a linear operator with a fixed size $N$, the triglobal one tends to be denser because of the larger stencils required to compute base flow derivatives in all spatial directions. When employing a direct linear system solver approach to compute the resolvent modes, special attention must be given to the LU factorization of the linear operator. Factorizing a triglobal resolvent operator results in higher fill-in ratios and increased memory requirements. Consequently, for triglobal resolvent analysis of high-Reynolds-number flows, numerical methods that address these cost-related matters are essential.

\subsection{Resolvent-guided flow control}
\label{sec:control}

A natural usage of the insights from resolvent analysis is the design of active and passive flow control strategies. Resolvent analysis elucidates the characteristics of optimal forcing input that can be amplified in the flow field and describes how the flow responds to actuation input.  The understanding of the input-output dynamics forms the basis for flow estimation and control design \cite{luhar2014opposition, martini2022resolvent, herrmann2023interpolatory}.  In the actual nonlinear flow, the level of amplified perturbations can exceed the linear regime and consequently modifies the base flow. However, describing what modification can be made to the base flows by the introduced forcing requires some extensions of basic resolvent analysis.  While the analysis identifies the forcing shape and frequency that can potentially result in an effective departure from the base state, full nonlinear flow simulations (CFD) and experiments are necessary to fully identify the modification achieved by the control inputs.  

\begin{figure}
\centering
\includegraphics[width=\textwidth]{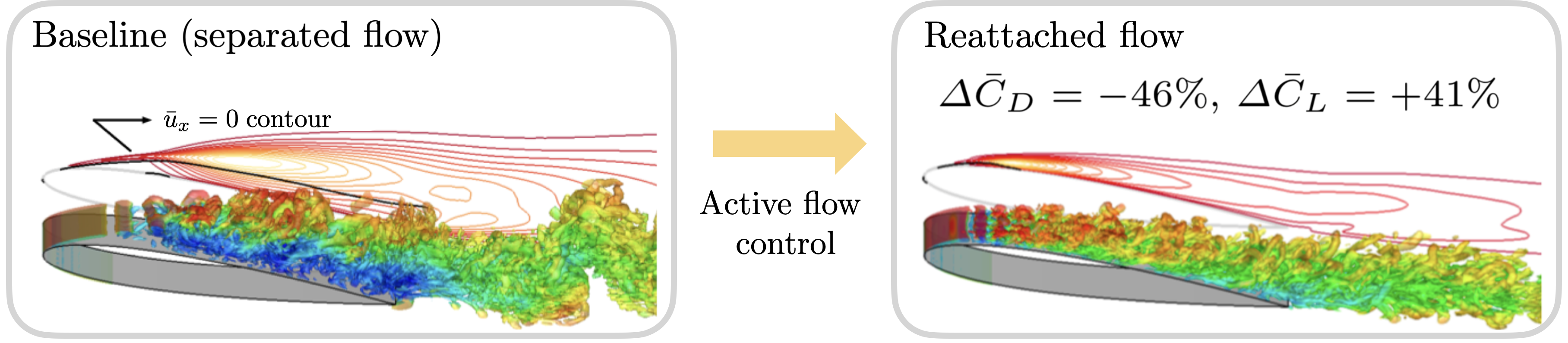}
\caption{\label{fig:AFC_NACA0012} An example of flow control guided by resolvent analysis.  Flow separation is suppressed using the frequencies and wavenumbers of high resolvent gain and insights from the modal Reynolds stress \cite{yeh2019resolvent}.}
\end{figure}

Flow control design using resolvent analysis has been demonstrated in a variety of applications, including turbulent channel flows \cite{toedtli2019predicting,nakashima2017assessment}, vortex shedding control around cylinders \cite{jin2020feedback,lin2023flow}, suppression of laminar separation bubbles over airfoils \cite{yeh2019resolvent,gross2024laminar}, attenuation of pressure fluctuations in supersonic turbulent cavity flow \cite{liu2021unsteady}, and control of separation and tip vortices around 3-D wings \cite{ribeiro2024control}.  In addition to relying on the resolvent gain as a quantitative metric to identify effective control parameters, it is also possible to leverage the resolvent mode structures.  Figure~\ref{fig:AFC_NACA0012} shows an example of exploiting both resolvent gain and modes to suppress flow separation over an airfoil \cite{yeh2019resolvent}.  Here, in addition to the resolvent gain, the modal Reynolds stress, given by $R_{ij} = \left\langle \hat{q}_i, \hat{q}_j\right\rangle$ (subscript $i$ denotes the $i$th velocity component of the response mode) is also used to quantify the level of momentum mixing enabled by the excited coherent structures.  Results show successful suppression of flow separation and improved aerodynamic performance with the resolvent-guided flow control design.  This approach significantly can narrow down the design candidates for effective flow control setups with low computational burden instead of performing an enormous number of large scale numerical computations for a full parametric investigation \cite{yeh2019resolvent}. 

\section{Concluding remarks}
\label{sec:Conclu}

We presented the resolvent analysis and offered a guide on its implementations and applications.  This paper offered details on the formulation and discussed subtleties in the foundations that should be kept in mind as one applies resolvent analysis to a chosen base flow.  Here, the presentations are offered for the compressible Navier--Stokes equations.  However, the general concepts that were discussed here can be applied to the incompressible Navier--Stokes equations and other dynamical systems.

Because the main purpose of this paper is to act as a guide and lower the barrier for new practitioners to resolvent analysis, we did not provide an extensive survey of the existing work.  We instead point readers to recent review articles on resolvent analysis \cite{Jovanovic:ARFM21} and modal analysis \cite{taira2017modal,taira2020modal} to gain additional insights and to observe a variety of applications.

As resolvent analysis is considered one of the operator-based methods, access to the discretized version of the linearized Navier--Stokes equations that generally come in the form of a CFD program or spatial discretization schemes is required. As such, grid generation and spatial discretization must be carefully performed.  Due to these reasons, there is no universal resolvent analysis computer program that works for all flows, in contrast to data-driven methods such as the proper orthogonal decomposition and dynamic mode decomposition that do not require access to the linearized Navier--Stokes operator.  However, to support our goal of making resolvent analysis a viable choice of examination tool for students and practitioners, this article attempted to provide as many necessary details, including pseudocodes, and know-how as possible to develop the resolvent analysis tool.  We hope that the readers find this as a useful guide when they delve into this powerful method.  

We also hope this article sparks the interest of those unfamiliar with resolvent analysis, encouraging them to explore its potential for uncovering valuable insights for their research endeavors.  Resolvent analysis has proven to be useful for examining unsteady flow physics, deriving reduced-order models, and developing flow control designs.  The resolvent analysis uncovers the dominant input-output dynamics of the flow with two sets of modes that can serve as the basis to capture the dominant flow oscillations.  The close relationship of resolvent analysis to dynamical systems and control theories also gives great potential for novel modeling and control techniques for turbulent flows.  We will be delighted if this article serves as an invitation for many new students and researchers to the world of resolvent analysis.

\section*{Acknowledgments}

The authors thank
Gabrielle Claus, 
Tim Colonius, 
Daniel Garmann, 
Vedasri Godavarthi, 
Dylan House,
Mihailo Jovanovi\'c,
Soshi Kawai, 
Yoimi Kojima, 
Qiong Liu, 
Beverley McKeon, 
Tulio Ricciardi, 
Peter Schmid, 
Calum Skene, 
Yiyang Sun, 
Vassilis Theofilis, and 
Yonghong Zhong
for the enlightening and delightful discussions on resolvent analysis.
We are also grateful to all of the students and postdocs in our groups who provided valuable feedback on the draft version of this manuscript.
KT acknowledge 
the US Army Research Office (W911NF-21-1-0060), 
the US Air Force Office of Scientific Research (FA9550-21-1-0174, FA9550-22-1-0013), and 
the US Department of Defense Vannevar Bush Faculty Fellowship Program (N00014-22-1-2798)
for supporting various aspects of the work reported in this paper.

\begin{appendices}

\section{The linearized Navier-Stokes equations}
\label{AppendixA}

In this appendix, we provide the full linearized compressible Navier-Stokes equations about a chosen base flow $\mathbf{q}_b=(\bar{\rho},\overline{\rho \mathbf{u}},\overline{\rho e})$.  The linearized mass, momentum, and energy conservation equations with source/forcing terms, respectively, are
\begin{align}
    \frac{\partial \rho^\prime}{\partial t}
    &+ \bnabla \cdot (\rho\mathbf{u})^\prime
    = f_\rho,\label{eq:LinearConservative1}
    \\
    \frac{\partial (\rho \mathbf{u})^\prime}{\partial t}
    &+ \bnabla \cdot \left[ (\rho \mathbf{u})^\prime\otimes \bar{\mathbf{u}}+\widebar{\rho \mathbf{u}}\otimes \mathbf{u}^\prime \right]
    \nonumber \\ 
    & \qquad = -\bnabla p^\prime+\bnabla \cdot \left[\mu\;(\bnabla\mathbf{u}^\prime+\bnabla\mathbf{u}^\prime{}^T)-\frac{2}{3}\mu\bnabla\cdot\mathbf{u}^\prime\:\mathrm{I}\right] 
    + \mathbf{f}_{\rho \mathbf{u}},\label{eq:LinearConservative2}
    \\
    \frac{\partial (\rho e)^\prime}{\partial t} 
    & +\bnabla \cdot \left[ \widebar{\rho \mathbf{u}}e^\prime+(\rho\mathbf{u})^\prime\bar{e}+\bar{p}\mathbf{u}^\prime+p^\prime\bar{\mathbf{u}} \right] \nonumber \\ 
    & \qquad =\bnabla \cdot (K\bnabla T^\prime)+\bnabla \cdot \left(\bar{\mathbf{u}}\;\left[\mu\;(\bnabla\mathbf{u}^\prime+\bnabla\mathbf{u}^\prime{}^T)-\frac{2}{3}\mu\bnabla\cdot\mathbf{u}^\prime\:\mathrm{I}\right]\right) \nonumber \\ 
    & \qquad \qquad +\bnabla \cdot \left(\mathbf{u}^\prime\;\left[\mu\;(\bnabla\bar{\mathbf{u}}+\bnabla\bar{\mathbf{u}}^T)-\frac{2}{3}\mu\bnabla\cdot\bar{\mathbf{u}}\:\mathrm{I}\right]\right) 
    + f_{\rho e}, 
    \label{eq:LinearConservative3}
\end{align}
where we have 
\begin{equation}
    \begin{split}
        \mathbf{u}^\prime &= \frac{(\rho \mathbf{u})^\prime }{\bar{\rho}}-\frac{\widebar{\rho \mathbf{u}}}{\bar{\rho}^2}\rho^\prime, \qquad 
        e^\prime = \frac{(\rho e)^\prime}{\bar{\rho}}-{\frac{\overline{\rho e}}{\bar{\rho }^2}}\rho^\prime,\\
        T^\prime &= \frac{e^\prime-\bar{\mathbf{u}}\cdot\mathbf{u}^\prime}{c_v}, \qquad 
        p^\prime = \rho^\prime R\bar{T}+\bar{\rho }R T^\prime.\\
    \end{split}\label{eq:linPrimitivevariable}
\end{equation}
Here, we have assumed that $\mu$ and $K$ are constant.  Note that the above set of equations is written in continuous form and requires spatial discretizations.

These equations can now be written in short-hand notation as
\begin{equation}
    \frac{d \mathbf{q}^\prime}{d t}=\mathcal{L}\mathbf{q}^\prime +\mathbf{f}^\prime,
\end{equation}
which is Eq.~\ref{eq:lin} in the main discussion.  The above formulation considers the resolvent analysis in a three-dimensional domain with three inhomogeneous (non-periodic) spatial directions, referred to as the triglobal analysis \cite{Theofilis:ARFM11}.  For implementation in primitive variables, details can be found in \cite{ohmichi2021matrix}.

When one of the spatial directions is homogeneous (periodic), then the above formulation reduces to a biglobal formulation which can be used when the base flow is two-dimensional $\mathbf{q}_b = \mathbf{q}_b(x,y)$.  In this case, perturbations can be expressed as 
\begin{equation}
    \mathbf{q}(x,y,z,t)=\mathbf{q}(x,y) e^{-i\beta z -i\omega t}    
\end{equation}
implying that ${\partial}/{\partial z}=-i\beta$, where $\beta$ denotes the spanwise wavenumber.   Therefore, the linearized Navier-Stokes operator becomes comprised of
\begin{equation}\label{eq:LinopSum}
    \mathcal{L}= \mathcal{L}_0 - i\beta \mathcal{L}_1 +\beta^2 \mathcal{L}_2,
\end{equation}
where $\mathcal{L}_0$ contains all linear terms not related to the derivative in the homogeneous direction and 
$\mathcal{L}_1$ and $\mathcal{L}_2$ collect all terms that have first and second-order derivatives in the homogeneous direction, respectively.
These three operators $\mathcal{L}_0$, $\mathcal{L}_1$, and $\mathcal{L}_2$ need to be computed only once because Eq.~(\ref{eq:LinopSum}) can be used to compose $\mathcal{L}$ for different values of $\beta$. 


\section{Pseudocodes}
\label{AppendixB}

The pseudocodes \ref{alg:ACompact} to \ref{alg:rsvd3} referenced in Sect.~\ref{sec:Methods} are provided below.

\begin{algorithm}
\caption{Compact matrix $\tilde{A}$}\label{alg:ACompact}
\begin{algorithmic}
\Require Linear operator (discretized) $\mathcal{L} \in \mathbb{C}^{N \times N}$, $m$ dimension of Krylov subspace
\Function{projected\underline{ }matrix}{$\mathcal{L}$,$m$}
\State $\mathbf{x}_1 \gets \textsf{randn}(N)$\Comment{Initial vector}
\State $\mathbf{x}_1 \gets \mathbf{x}_1/||\mathbf{x}_1||_2$
\State $V_1=[\mathbf{x}_1]$
\State $\tilde{A}=[\;]$
 \For{$j=1:m-1$} 
 \State $\mathbf{tmp} \gets [-i\omega I-\mathcal{L}]  \symbol{92} \mathbf{x}_j$ \Comment{Solve direct system}
 \State $\mathbf{y} \gets [-i\omega I-\mathcal{L}]^*\symbol{92} \mathbf{tmp}$\Comment{Solve adjoint system}
 \State $\mathbf{a}=V_j^T\mathbf{y}$ \Comment{Projection onto $V_j$}
 \State $\mathbf{y}=\mathbf{y}-V_j\mathbf{a}$ \Comment{Orthognalization}
\If{$j=1$}
 \State $\tilde{A}\gets\begin{bmatrix}
    \mathbf{a}\\
     ||\mathbf{y}||_2
 \end{bmatrix}$
  \Else
   \State $\tilde{A}\gets\begin{bmatrix}
     \tilde{A} &\mathbf{a}\\
     0 & ||\mathbf{y}||_2
 \end{bmatrix}$
  \EndIf{}
 \State $\mathbf{x}_{j+1} \gets \mathbf{y}/||\mathbf{y}||_2$
  \State $V_{j+1} \gets \begin{bmatrix}
      V_j & \mathbf{x}_{j+1} \end{bmatrix}$
\EndFor
 \State $\mathbf{tmp} \gets [-i\omega I-\mathcal{L}]  \symbol{92} \mathbf{x}_m$ 
 \State $\mathbf{y} \gets [-i\omega I-\mathcal{L}]^*\symbol{92} \mathbf{tmp}$
 \State $\mathbf{a}=V_m^T\mathbf{y}$ 
 \State $\tilde{A}\gets\begin{bmatrix}
     \tilde{A} &\mathbf{a}
 \end{bmatrix}$
\State \Return ($\tilde{A}$,$V_m$)
\EndFunction
\end{algorithmic}
\end{algorithm}

\begin{algorithm}
\caption{Compute resolvent modes from $\tilde{A}$}\label{alg:EVATilde}
\begin{algorithmic} 
\Require Linear operator (discretized) $\mathcal{L} \in \mathbb{C}^{N \times N}$, low-rank projection $\tilde{A}$ and orthogonalized Krylov subspace $V_m$
\Function{compute\underline{ }singular\underline{ }triplet}{$\mathcal{L}$,$\tilde{A}$,$V_m$}
\State  $[\sim, \tilde{F}] \gets \textsf{eig}(\tilde{A})$ \Comment{Eigenvalue decomposition of $\tilde{A}$}  
\State ${F} \gets V_m\Tilde{F}$  \Comment{Retrieving response modes}
\State $ \Tilde{Q} \gets [-i\omega I-\mathcal{L}] \symbol{92} {F}^*$ \Comment{Solve direct linear system to obtain $\Tilde{Q} \in \mathbb{C}^{N \times m}$}
\For{$j \gets 1$ to $m$}
    \State $\Sigma_{j,j} \gets \textsf{norm}(\Tilde{Q}_{[:,j]},2)$ \Comment{Columnwise norm of $\textsf{norm}(\Tilde{Q})$ retrieves gains}
    \State ${Q}_{:,j} \gets \Tilde{Q}_{[:,j]} / \Sigma_{j,j}$ \Comment{Columnwise normalization of $\Tilde{Q}$ retrieves response modes}
\EndFor
\State \Return (${Q},\Sigma, F$) 
\EndFunction
\end{algorithmic}
\end{algorithm}

\begin{algorithm}
\caption{Randomized $\mathcal{Q}B$ decomposition}\label{alg:qb}
\begin{algorithmic}
\Require Linear operator (discretized) $\mathcal{L} \in \mathbb{C}^{N \times N}$ 
\Function{randomized\underline{ }qb\underline{ }decomposition}{$\omega$,m}
\State $\Theta \gets \textsf{randn}(N,m)$
\State $Y \gets [-i\omega I-\mathcal{L}] \symbol{92} \Theta$ \Comment{Solve direct system to obtain $Y \in \mathbb{C}^{N \times m}$}
\State $\mathcal{Q} \gets \textsf{orthogonalize}(Y)$ \Comment{Orthogonalize to create sketch basis $\mathcal{Q} \in \mathbb{C}^{N \times m}$}
\State $B^* \gets [-i\omega I-\mathcal{L}^*] \symbol{92} \mathcal{Q}^*$ \Comment{Solve adjoint system to obtain $B \in \mathbb{C}^{N \times m}$}
\State \Return ($\mathcal{Q}$,B)
\EndFunction
\end{algorithmic}
\end{algorithm}

\begin{algorithm}
\caption{Compute resolvent modes: method 1}\label{alg:rsvd1}
\begin{algorithmic} 
\Require Sketch basis $\mathcal{Q}$ and low-rank projection $B$ of resolvent operator 
\Function{randomized\underline{ }resolvent\underline{ }opt1}{$\mathcal{Q}$,B}
\State  $[\Tilde{Q},\Sigma, F] \gets \textsf{svd}(B^*)$ \Comment{SVD of $B^*$ retrieves forcing modes}  
\State ${Q} \gets \mathcal{Q}\Tilde{Q}$  \Comment{Matrix multiplication retrieves response modes}
\State \Return (${Q},\Sigma, F$) 
\EndFunction
\end{algorithmic}
\end{algorithm}

\begin{algorithm}
\caption{Compute resolvent modes: method 2}\label{alg:rsvd2}
\begin{algorithmic} 
\Require Linear operator (discretized) $\mathcal{L} \in \mathbb{C}^{N \times N}$ 
\Function{randomized\underline{ }resolvent\underline{ }opt2}{$\omega$,B}
\State  $[\sim,\sim, F] \gets \textsf{svd}(B)$ \Comment{SVD of $B^*$ retrieves forcing modes}
\State $ \Tilde{Q} \gets [-i\omega I-\mathcal{L}] \symbol{92} {F}^*$ \Comment{Solve direct linear system to obtain $\Tilde{Q} \in \mathbb{C}^{N \times m}$}
\For{$j \gets 1$ to $m$}
    \State $\Sigma_{j,j} \gets \textsf{norm}(\Tilde{Q}_{[:,j]},2)$ \Comment{Columnwise norm of $\textsf{norm}(\Tilde{Q})$ retrieves gains}
    \State ${Q}_{:,j} \gets \Tilde{Q}_{[:,j]} / \Sigma_{j,j}$ \Comment{Columnwise normalization of $\Tilde{Q}$ retrieves response modes}
\EndFor
\State \Return (${Q},\Sigma, F$)
\EndFunction
\end{algorithmic}
\end{algorithm}

\begin{algorithm}
\caption{Compute resolvent modes: method 3}\label{alg:rsvd3}
\begin{algorithmic} 
\Require Linear operator (discretized) $\mathcal{L} \in \mathbb{C}^{N \times N}$ 
\Function{randomized\underline{ }resolvent\underline{ }opt3}{$\omega$,B}
\State  $[\sim,\sim, \Tilde{F}] \gets \textsf{svd}(B)$ \Comment{SVD of $B^*$ retrieves $\Tilde{F}$}
\State $ \Tilde{Q} \gets [-i\omega I-\mathcal{L}] \symbol{92} \Tilde{F}^*$ \Comment{Solve direct linear system to obtain $\Tilde{Q} \in \mathbb{C}^{m \times k}$}
\State $[Q,\Sigma, \Hat{F}] \gets \textsf{svd}(\Tilde{Q})$ \Comment{SVD of $\Tilde{Q}$ retrieves gains and response modes}
\State $F \gets \Tilde{F}\Hat{F}$ \Comment{Matrix multiplication retrieves forcing modes}
\State \Return (${Q},\Sigma, F$)
\EndFunction
\end{algorithmic}
\end{algorithm}

\end{appendices}

\newpage
\bibliography{sn-article}

\end{document}